\documentstyle[12pt,epsfig,multirow]{article}

\newcommand{\be}{\begin{eqnarray}}
\newcommand{\ee}{\end{eqnarray}}

\def\nue{{\nu_e}}
\def\anue{{\bar\nu_e}}
\def\numu{{\nu_{\mu}}}
\def\anumu{{\bar\nu_{\mu}}}

\newcommand{\chr}{\mbox{$\breve{\rm C}$erenkov~}}

\newcommand{\ms}{\Delta m^2_{21}}
\newcommand{\ma}{\Delta m^2_{31}}

\newcommand{\sss}{\sin^2 \theta_{12}}

\newcommand{\stch}{\sin^2 2\theta_{13}}

\newcommand{\sta}{\sin^22 \theta_{23}}

\newcommand{\sgnma}{\mathrm{sgn}(\Delta m^2_{31})}

\newcommand{\stcht}{\sin^2 2\theta_{13}{\mbox {~(true)}}}

\newcommand{\dcpt}{\delta_{\mathrm{CP}}{\mbox {~(true)}}}
\newcommand{\dcp}{\delta_{\mathrm{CP}}}

\newcommand{\sig}{$3\sigma$}

\newcommand{\br}{$^8$B~}
\newcommand{\li}{$^8$Li~}
\newcommand{\he}{$^6$He~}
\newcommand{\neon}{$^{18}$Ne~}

\def\gtap{\ \raisebox{-.4ex}{\rlap{$\sim$}} \raisebox{.4ex}{$>$}\ }

\newcommand{\ie}{{\it i.e.}}

\newcommand{\cf}{{\it c.f.}}
\newcommand{\etc}{{\it etc.}}
\newcommand{\eq}{Eq.}

\newcommand{\fig}{Fig.}

\newcommand{\Ref}{Ref.}
\newcommand{\Refs}{Refs.}
\newcommand{\Sec}{Section}

\newcommand{\stheta}{\sin^22\theta_{13}}
\newcommand{\deltacp}{\delta_{\mathrm{CP}}}
\newcommand{\ldm}{\Delta m_{31}^2}

\newcommand{\equ}[1]{\eq~(\ref{equ:#1})}
\newcommand{\figu}[1]{\fig~\ref{fig:#1}}

\begin{document}

\thispagestyle{empty}
\begin{flushright}
\texttt{HRI-P-08-02-001}\\
\texttt{CU-PHYSICS/03-2008}\\
\end{flushright}
\bigskip

\begin{center}
{\Large \bf Optimizing the greenfield Beta-beam} 

\vspace{.5in}

{\bf {Sanjib Kumar Agarwalla$^{\star,\dagger,a}$, 
Sandhya Choubey$^{\star,b}$,\\ Amitava Raychaudhuri$^{\star,\dagger,c}$, 
Walter Winter$^{\sharp,d}$}}
\vskip .5cm
$^\star${\normalsize \it Harish-Chandra Research Institute,} \\
{\normalsize \it Chhatnag Road, Jhunsi, Allahabad  211019, India}\\
\vskip 0.4cm
$^\dagger${\normalsize \it Department of Physics, University of Calcutta,} \\ 
{\normalsize \it 92 Acharya Prafulla Chandra Road, Kolkata  700009, India}\\
\vskip 0.4cm
$^\sharp${\normalsize \it Institut f{\"u}r Theoretische Physik und
  Astrophysik, Universit{\"a}t W{\"u}rzburg,}\\
{\normalsize \it Am Hubland, D-97074 W{\"u}rzburg, Germany}
\vskip 1cm

{\bf ABSTRACT}
\end{center}

\noindent
We perform a comprehensive and detailed 
comparison of the physics reach of Beta-beam 
neutrino experiments between two pairs of plausible 
source ions, ($^8$B, $^8$Li) and ($^{18}$Ne, $^6$He).
We study the optimal choices for the 
baseline, boost factor, and luminosity. 
We take a 50 kton iron calorimeter, {\it a la} 
ICAL@INO, as the far detector. We follow two 
complementary approaches for our study: (i) Fixing 
the number of useful ion decays and boost factor of the beam, and 
optimizing for the sensitivity reach between the two pairs of ions
as a function of the baseline. 
(ii) Matching the shape of the spectrum between the
two pairs of ions, and studying the requirements for baseline,
boost factor, and luminosity. We find that for each pair
of ions there are two baselines with very good sensitivity reaches:
a short baseline with $L \, [\mathrm{km}]/
\gamma \simeq 2.6$ ($^8$B+$^8$Li) and 
$L \, [\mathrm{km}]/\gamma \simeq 0.8$  
($^{18}$Ne+$^6$He), and a long ``magic'' baseline.
For $\gamma \sim 500$, one would optimally use $^{18}$Ne and 
$^6$He  at the short baseline for CP violation, $^8$B and 
$^8$Li at the magic baseline for the mass hierarchy, and either  
$^{18}$Ne and $^6$He at the short baseline or $^8$B and $^8$Li at 
the magic baseline for the $\stheta$ discovery. 

\vfill

\noindent $^a$ email: sanjib@hri.res.in

\noindent $^b$ email: sandhya@hri.res.in

\noindent $^c$ email: raychaud@hri.res.in

\noindent $^d$ email: winter@physik.uni-wuerzburg.de

\newpage

\section{Introduction}

Spectacular results from a series of neutrino oscillation 
experiments over the last four decades~\cite{solar,kl,atm,
chooz,k2k,minos,limits}
have paved the way for the ``golden'' age of neutrino physics. 
Determining the hitherto unknown mixing angle $\theta_{13}$, 
the CP phase $\deltacp$, and the sign of $\ldm$, \ie, 
the neutrino mass hierarchy\footnote{Though we call this 
the neutrino mass {\it hierarchy}, 
what we 
mean is basically the {\it ordering} of the neutrino 
mass states. Therefore, our discussions are valid for 
both hierarchical as well as quasi-degenerate mass spectra. 
We define $\Delta m_{ij}^2 = m_i^2 - m_j^2$
and refer to
$sgn(\ma)$ as the neutrino mass hierarchy --
$sgn(\ma) >0$ is called ``normal hierarchy''(NH)
while $sgn(\ma) <0$ is called ``inverted hierarchy''(IH).
},
have emerged as the next frontier in this field. 
All these three quantities can be probed by 
experimentally measuring the so-called ``golden'' channel~\cite{golden}
oscillation probability $P_{e\mu}$ (or its T-conjugate 
channel $P_{\mu e}$). 
A series of ambitious projects are under discussion 
which plan to use this oscillation channel. 
The on-going and near future experiments include 
the MINOS experiment in the US~\cite{minos}, and the CNGS experiments 
ICARUS~\cite{icarus} and OPERA~\cite{opera} in Europe.
Next experiments in line will be T2K in Japan~\cite{t2k}
and NO$\nu$A in US \cite{nova}. All these experiments 
will use muon neutrino beams from conventional accelerator 
sources in order to observe $P_{\mu e}$. 
Collectively and in combination with short-baseline reactor
experiments, such as Double Chooz \cite{chooz2}, these experiments 
are expected to  improve the bound on $\theta_{13}$ to about 
$\stch < 0.01$  ($90\%$ CL)~\cite{huber10}.
The mass hierarchy and CP violation, 
though in principle accessible using the combined 
data from the T2K and NO$\nu$A experiments, can be determined 
only for values of $\stheta$~(true) close to the current
bound and for some fraction of the possible values of 
the CP phase $\deltacp$~(true).\footnote{We distinguish between the 
``true'' values of the oscillation parameters, which are 
the values chosen by Nature, and their fitted values. 
Throughout this paper we 
denote the true value of a parameter
by putting ``(true)'' after the symbol for the parameter.}
The sensitivity of these experiments is mainly restricted
by statistics, while for larger luminosity set-ups,
the intrinsic 
$\nue$ background poses a natural limitation for
experiments sensitive to 
$\numu$ oscillations into $\nue$. 
Therefore, if Nature has not been very kind
we will need larger experiments to complete our understanding 
of the neutrinos, possibly using an alternate technology.

In order to access small values of $\stheta$, there are several requirements:
One needs to 
have low backgrounds, more statistics, and 
reduced systematical uncertainties.
As far as the low background requirement is concerned, it 
is an advantage to use a pure flavor neutrino beam 
without any intrinsic beam contamination. 
One such approach has been proposed by Piero Zucchelli~\cite{zucc}: 
Radioactive nuclides are created by impinging a target 
by accelerated protons. These unstable nuclides are 
collected, fully ionized, 
bunched, accelerated and then stored in a 
decay ring (see for {\it e.g.} \cite{lindroos,betabeampage}).  
The decay of these highly boosted ions 
in the straight sections of the decay ring produces the 
so-called Beta-beam. 
An alternative approach
is the so-called Neutrino Factory (NuFact) \cite{geer}.
It involves  producing, collecting, cooling, 
accelerating, and circulating 
muon packets in a storage ring. The decay of 
accelerated muons (antimuons) 
in the straight sections of the storage 
rings produce $\numu$ and $\anue$ ($\anumu$ and $\nue$) beams. 
The presence of the $\anue$ ($\nue$) in the beam allows for 
the observation of the  $P_{\bar e\bar \mu}$ ($P_{e\mu}$)
oscillation probability in the 
far detector. Since both $\numu$ (or $\anumu$) 
from the original beam as well as $\anumu$ (or $\numu$) 
from the oscillated $\anue$ (or $\nue$) will be 
arriving at the detector, it must have the ability 
to distinguish one from the other. The most accepted 
candidate is the magnetized iron detector, though there are 
several proposals with more expensive and elaborate designs
and therefore better performance~\cite{mind}. 
Statistics can be increased by a higher beam power and the size and 
efficiency of the detector. Beam-related 
systematic uncertainties can be reduced to a large extent 
by working with a two detector set-up, one very close to the beam 
line and another serving as the far detector. The systematic 
uncertainties coming from the lack of knowledge of the 
neutrino-nucleus interaction 
cross-sections are another important source of error. 
These can be controlled to some degree by the near-far two detector 
set-up, but they cannot be canceled completely~\cite{crossthomas}.
Beam-related backgrounds are extremely small for the 
NuFact and Beta-beam experiments because they either use leptonic
decays (NuFact) or a flavor-pure beam (Beta-beam). The detector 
backgrounds coming mainly from 
neutral current interactions and 
mis-identification of particles, can be reduced by imposing 
intelligent cuts. The atmospheric neutrino backgrounds, 
which can be important for Beta-beams at lower energies,   
can be suppressed using timing and directional information. 

The performance and physics reach of these expensive and 
ambitious experiments have been the subject of much 
discussion for the past few years~\cite{iss}. There has been a  
plethora of papers on this 
issue\footnote{A summary of the potential of 
selected NuFact and Beta-beam set-ups 
have been compiled by the physics working group of the 
International Scoping Study for a future Neutrino Factory, 
Superbeam and Beta-beam, in their report~\cite{issphysics}.}, 
most of which have addressed the problem of ``parameter degeneracies''.
Even if both neutrinos and antineutrinos are used, there are
three types of discrete degeneracies in the golden channel: 
\begin{enumerate}
\item the ($\theta_{13},\deltacp$) intrinsic degeneracy~\cite{intrinsic},
\item the ($\sgnma,\deltacp$) degeneracy~\cite{minadeg},
\item the ($\theta_{23},\pi/2-\theta_{23}$) degeneracy~\cite{th23octant}.
\end{enumerate}
Together they can result in up to eight-fold degenerate 
solutions~\cite{eight}, 
severely deteriorating the sensitivity of the 
experiment. The variety of suggestions to solve
this problem includes
combining data from several experiments observing 
the golden channel, but with different baselines $L$ 
and neutrino energies $E$~\cite{intrinsic,diffLnE,t2ksimulation},
combining data from accelerator experiments observing 
different oscillation channels~\cite{silver,dissappear,pee},
combining the golden channel data with those 
from atmospheric neutrino ~\cite{addatm,cernmemphys} or reactor 
antineutrino experiments \cite{addreact}. A particularly 
attractive way of completely resolving at least two of 
the three degeneracies is to perform the experiment 
at the  ``magic baseline''~\cite{magic,magic2,petcov}. 
This magic baseline reflects the 
characteristic oscillation wavelength corresponding to Earth matter. 
One can show that at this baseline, for reasonably small values 
of $\theta_{13}$, the $\deltacp$-dependent terms 
vanish. The $\deltacp$-dependence is therefore 
reduced and one can eliminate the ($\theta_{13},\deltacp$)
and ($\sgnma,\deltacp$) degeneracies, resulting in 
tremendous sensitivity to $\theta_{13}$ and 
$\sgnma$. The sensitivity reach of a NuFact
experiment at the magic baseline can be found in 
\cite{magic,nufactoptim}. 
The idea for a magic baseline 
Beta-beam experiment with a similar performance for $\theta_{13}$ and 
$\sgnma$ was put forth in 
\Refs~\cite{paper1,betaino1,betaino2}. 

The sensitivity reach of an experiment depends crucially on 
beam, baseline, and detector properties. It is therefore
important to ask which 
beam, baseline and detector set-up would qualify as the  
{\it optimal} choice in order to obtain the best $\stheta$ reaches 
for the three quantities we have set out to measure, \ie, 
$\theta_{13}$, $\sgnma$, and $\deltacp$. 
For the NuFact, this 
detailed exercise 
was performed in \Ref~\cite{nufactoptim}. The sensitivity to 
each of the 
three parameters mentioned above was studied 
as a function of the baseline and muon neutrino energy. 
It was demonstrated that the minimal muon neutrino 
energy acceptable for the magnetized iron detector was about $20$~GeV. 
The optimal baseline for 
probing $\theta_{13}$ and $\sgnma$ is the 
magic baseline, whereas best sensitivity to CP violation 
is expected at $L\sim 3000$ to $5000$~km~\cite{nufactoptim}. 

For Beta-beams, a variety of 
plausible set-ups have been proposed in the literature~\cite{cernmemphys,
paper1,betaino1,betaino2,oldpapers,donini130,
doninibeta,newdonini,bc,bc2,fnal,betaoptim,volpe,doninialter,rparity,boulby}.
The proposal which poses minimal  challenge 
for the Beta-beam design, is the 
commonly called CERN-MEMPHYS project~\cite{cernmemphys,oldpapers,donini130}. 
It proposes to use the EURISOL ion source to produce the radioactive 
source ions $^{18}$Ne  
and $^6$He, and demands a Lorentz boost factor $\gamma \simeq 100$ for them, 
which can be produced using the existing
accelerator facilities at CERN. The far detector MEMPHYS, 
a megaton water detector with fiducial 
mass of 440~kton, will have to be built in the 
Fr\'ejus tunnel, at a distance of 130 km from CERN. 
Another possible Beta-beam set-up 
using water detector but higher boost factors and 
an intermediate baseline option was put forth in 
\cite{bc,bc2} (see also \cite{doninialter}). In these papers 
authors have used a high $\gamma$ $^{18}$Ne  
and $^6$He 
Beta-beam option at CERN and 
440 kton fiducial volume water detector at 
GranSasso or Canfranc, which corresponds to $L=730$ and 650 km 
respectively. Excellent sensitivity to $\theta_{13}$ and 
CP violation is expected \cite{bc,bc2} from this proposal.  
Another high performance set-up proposed 
in~\cite{paper1,betaino1,betaino2} would use a 
high $\gamma$ Beta-beam and a magnetized iron detector  
placed at a distance close to the magic baseline. Since 
high energy neutrinos are mandatory for 
achieving near-resonant matter effects required for 
the desired performance of this set-up, 
one needs to employ an alternative set of source ions, 
$^8$B and $^8$Li~\cite{rubbia,mori}. The end-point 
energy of these ions are larger than those of \neon and \he 
by factor of about 3.5, and hence optimal neutrino energies can be 
obtained with a $\gamma$ between 
350 and 650. 
As one possible option, the Beta-beam could be
targeted from CERN 
towards the India-Based Neutrino Observatory (INO)~\cite{ino}.
The CERN to INO distance corresponds to 7152 km, which is 
almost magic. Therefore, this experiment yields sensitivity to 
$\theta_{13}$ and $\sgnma$, both of 
which could be outperformed only by the 
NuFact experiment at the magic baseline distance. 
Set-ups with a neutrino beam from CERN to GranSasso or
CanFranc~\cite{doninibeta}, from CERN to Boulby mine 
\cite{boulby}  
and from Fermilab~\cite{fnal} ($L \sim
300$ km) have also been proposed, 
and their sensitivity reach has
been explored. Set-up with two sets of source ions 
with different boost factor for 
each set but with the same baseline was proposed in 
\cite{doninialter}. In \cite{newdonini} the authors 
consider the complementary situation where they take 
only one set of source ions, \br and $^8$Li, with $\gamma=350$ 
and two different baselines, $L=2000$ km and 7000 km. 
Very high gamma Beta-beam options have been studied in
\Refs~\cite{paper1,bc2,betaoptim}. The physics potential 
of low energy Beta-beam option was probed in \cite{volpelow}. 
A comparison of the physics reach among different
Beta-beam experimental proposals can be found in
\Ref~\cite{volpe,nf07sc}.

In contrast to earlier works, we
study the baseline optimization as a function of the ion
pair used $^{18}$Ne+$^6$He  or $^8$B+$^8$Li, and we discuss the
impact of the luminosity. In addition, we perform a simultaneous
optimization of $L$ and $\gamma$. All comparisons are performed
for the same detector, which is a 50 kton magnetized iron
calorimeter. Note that the magnetization of the detector, which
is mandatory for the Neutrino Factory, is only used for a
reduction of the backgrounds.  We study the performance with
respect to $\theta_{13}$, $\sgnma$, and CP violation.  
paper is organized as follows. We describe the Beta-beam
experiment and our analysis procedure in \Sec~2. In \Sec~3, we
optimize the baseline $L$ of the experiment for fixed sets of
$\gamma$'s. In \Sec~4, we then perform a  simultaneous
optimization over $L$ and $\gamma$, and we discuss the
requirements for a similar beam spectrum.  In \Sec~5, we then
show the impact of the Beta-beam luminosity, and determine the
best combination of $L$, $\gamma$, and $N_\beta$ (the number of
useful ion decays per year). Our  conclusions can be found in
\Sec~6.

\section{Simulation of the Beta-beam Experiment}

Here we describe the experimental set-up for our 
proposed  Beta-beam  facility. We give the details of the flux and 
detector set-up we have used in our analysis.

\subsection{The Flux}
\label{sec:flux}

\begin{table}[t]
\begin{center}
\begin{tabular}{|c|c|c|c|c|c|} \hline
   Ion & $\tau$ (s) &
$E_0$ (MeV)
   & $f$& Decay fraction & Beam \\
\hline
  $^{18} _{10}$Ne &   2.41 & 3.92&820.37&92.1\%& $\nu_{e}$    \\
  $^6 _2$He   &   1.17 & 4.02&934.53&100\% &$\bar\nu_{e}$    \\
\hline
 $^{8} _5$B& 1.11 & 14.43&600872.07&100\%&$\nu_{e}$    \\
 $^8 _3$Li& 1.20 &13.47 &425355.16& 100\% & $\bar\nu_{e}$    \\
\hline
\end{tabular}
\caption{\label{tab:ions}
Beta decay parameters: lifetime $\tau$, 
electron total end-point energy 
$E_0$, $f$-value
and decay fraction for various ions~\cite{beta}. }
\end{center}
\end{table}

\begin{figure}[t]
\includegraphics[width=0.49\textwidth]{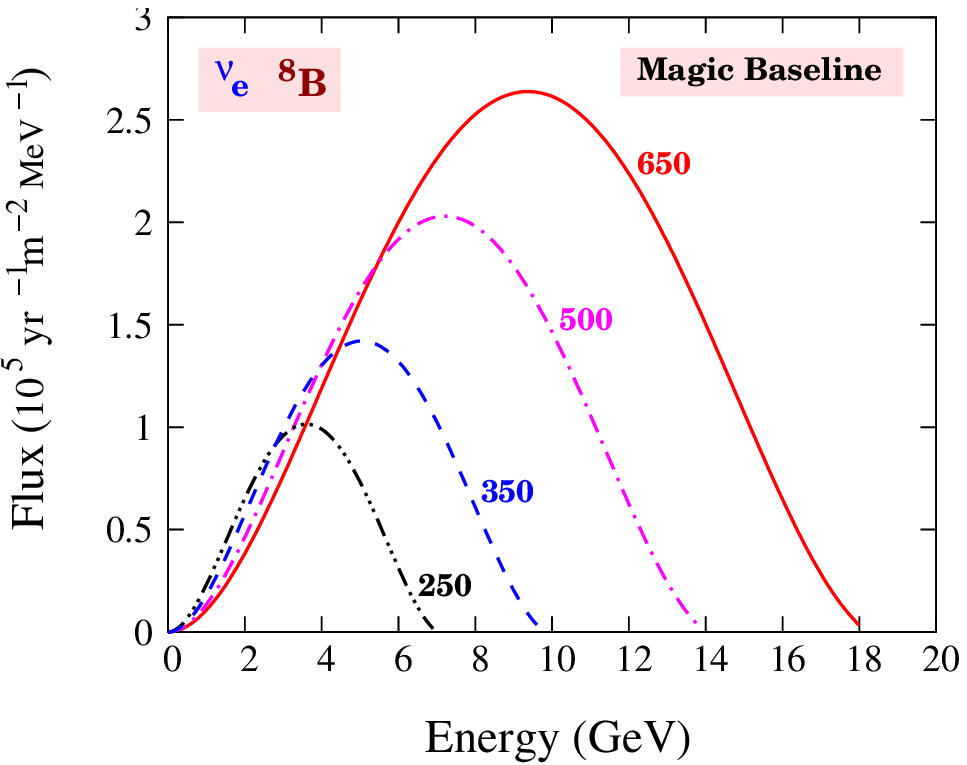}
\includegraphics[width=0.49\textwidth]{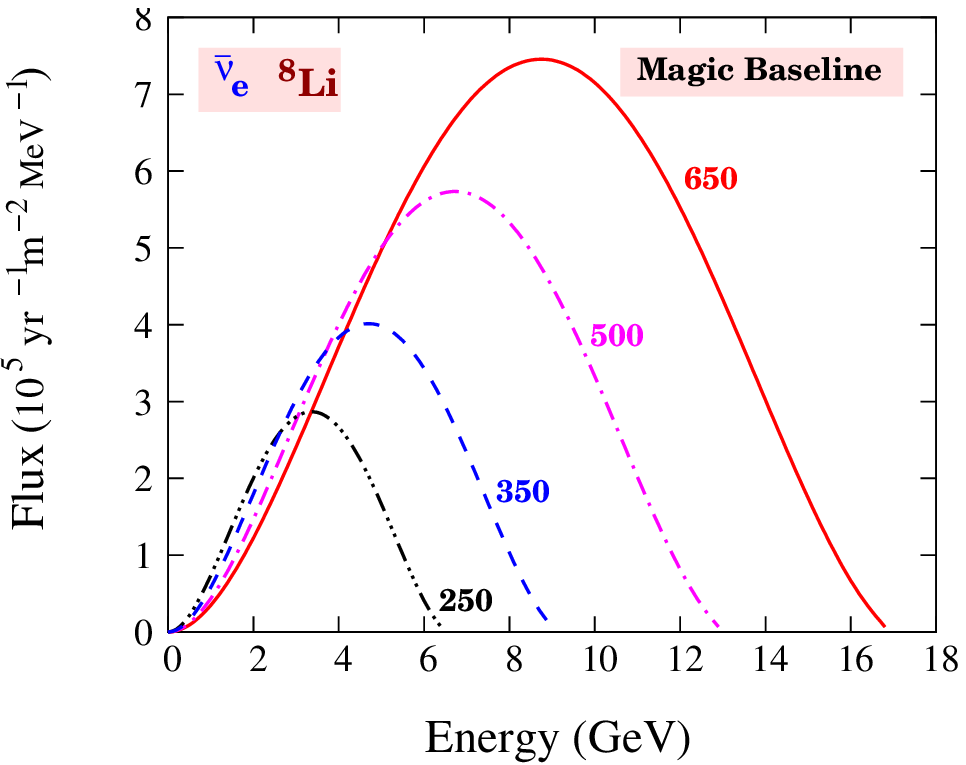}
\includegraphics[width=0.49\textwidth]{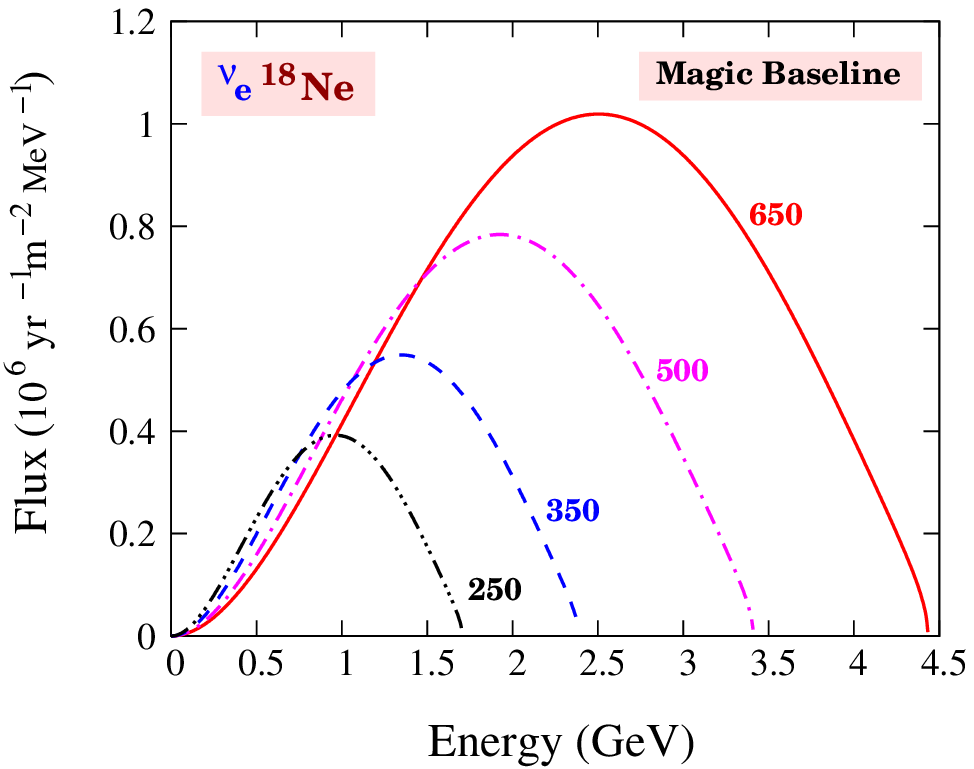}
\includegraphics[width=0.49\textwidth]{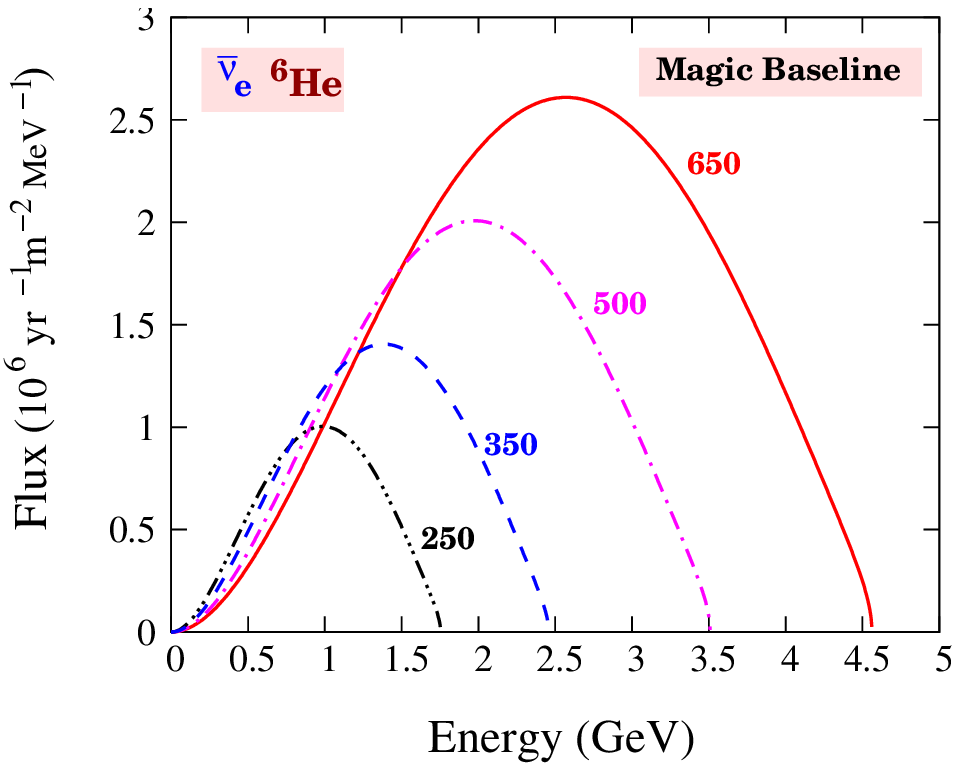}
\caption{\label{fig:flux}
The unoscillated 
Beta-beam flux spectrum arriving at a detector 
placed at the magic baseline. The upper panels are 
for \br (left panel) and \li (right panel), 
while the lower panels are for \neon (left panel) and \he (right panel).
}
\end{figure}

A Beta-beam~\cite{zucc} is an intense and highly 
collimated source of pure $\nue$ or $\anue$ flux, 
produced from the decay of beta unstable radioactive ions.
These unstable ions are created by impinging 
a target with high energy protons. Subsequently, the 
ions are collected, bunched, accelerated and 
stored in a decay ring. The standard design 
of the decay ring comprises of a racetrack shaped 
tunnel. When the ions decay along the straight sections, 
they produce a  $\nue$ or $\anue$ beam. 

This neutrino beam would be very suitable for precision 
experiments because it is  mono-flavor 
and hence beam related backgrounds are almost 
absent. The neutrino spectrum depends only on the beta decay total 
end-point energy $E_0$ and the Lorentz boost of the radioactive ions $\gamma$. 
The spectral shape can therefore be very well determined. 
The flux normalization is given directly by $N_\beta$, 
the number of 
useful ion decays per year in the straight section of the 
storage ring. 
The standard  numbers taken for the \neon and \he are 
$1.1 \times 10^{18}$ ($\nu_e$) and $2.9\times 10^{18}$ 
($\bar{\nu}_e$) useful 
decays per year, respectively \cite{beamnorm}.
Wherever not explicitly mentioned, these reference numbers of
useful ion decays for $\nu_e$ and $\bar{\nu}_e$ are chosen.
Note, however, that new ideas suggest
luminosities higher even by a factor of ten or so, depending on
the isotopes used,   by using a 
recirculating ring to improve the performance of
the ion source \cite{rubbia,mori}. The total luminosity
is given by the product of useful ion decays/year $\times$
running time $\times$ detector mass $\times$ detection efficiency.  
Throughout the study we 
will consider  five years of neutrino and 
five years of antineutrino running.  
Since the reference luminosity might not be reachable 
for different reasons, or it may be much higher because of
a better ion source, larger detector, \etc, we will include it 
as a parameter in this study.

The beam divergence is controlled by the Lorentz boost $\gamma$.
Hence by increasing $\gamma$, we can produce a higher beam 
collimation and increase the beam intensity along the 
forward direction $\propto \gamma^2$. However, note that though the 
intensity can be increased by 
choosing either a higher $N_\beta$ or $\gamma$, they might produce 
very different neutrino beams. While increasing 
$N_\beta$ merely increases the overall normalization 
of the flux, increasing $\gamma$ increases both the 
total flux as well as the average energy of the beam. 
This dependence of the beam flux on $\gamma$ is illustrated in \figu{flux}.
Notice that we assume the same 
$\gamma$ for both neutrino as well as antineutrino 
modes. 

Another crucial aspect associated with Beta-beams is 
the choice of the beta unstable ion. 
The properties that a suitable ion should have 
include a high production yield, large 
decay fraction, reasonably long lifetime, and 
preferably lower $Z/A$ ratio. The most widely discussed ions are 
$^{18}$Ne and $^6$He, 
which would produce a $\nue$ and $\anue$ beam 
respectively. The details of the source ions for Beta-beams
are given in Table~\ref{tab:ions}. 
The other pair of ions which have been proposed as 
an alternative to $^{18}$Ne and $^6$He, are 
$^8$B and $^8$Li~\cite{rubbia,mori}.
The main difference between $^8$B+$^8$Li compared to 
$^{18}$Ne+$^6$He is the higher end-point energy (see
Table~\ref{tab:ions}). The factor $\sim 3.68$ (3.35) difference
in end-point energy ensures that for the same peak
(anti)neutrino energy, approximately given by $\gamma E_0$, the
$\gamma$ required for \br (\li) will be 3.68 (3.35) times smaller
than that needed for $^{18}$Ne ($^6$He).  Since we assume the
same $\gamma$ for both ions within each pair, we use the average
difference in the total 
end-point energy $(3.35+3.68)/2 \simeq 3.5$ to
estimate the effects of $\gamma$.  The isotope dependence can be
also seen in \figu{flux}. The figure shows that 
the peak energy is approximately given
by $\gamma E_0$.

Let us discuss now the conditions to obtain a similar spectrum
(including normalization) when using different isotope pairs
at the same baseline. 
The purpose of this exercise is to derive the conditions
under which we produce matching neutrino energies and fluxes, 
and therefore deal with the same physics (including neutrino energies,
statistics, and, especially, matter effects).
If we neglect effects of the endpoint in the 
beta beam spectrum
(\ie, $E_0 \gg m_e$), 
we know from the beta beam flux
formula that the peak energy of the spectrum is approximately 
given by $\gamma E_0$,
and the total flux is proportional to $N_\beta \gamma^2$. 
In order to obtain a
spectrum with the {\it same peak energy and 
normalization} for two different
isotopes with very different endpoint energies (such as $^8$B and $^{18}$Ne),
we therefore have the following conditions (modulo endpoint effects):
\begin{equation}
\frac{N_\beta^{(1)}}{N_\beta^{(2)}} \simeq \left( \frac{E_0^{(1)}}
{E_0^{(2)}}\right)^2 \, , \quad 
\frac{\gamma^{(1)}}{\gamma^{(2)}} \simeq \frac{E_0^{(2)}}{E_0^{(1)}}
\label{equ:cond}
\end{equation}
From these matching conditions, one obtains the ratio of 
$N_\beta^{(1)}/N_\beta^{(2)}$ and $\gamma^{(1)}/\gamma^{(2)}$ 
needed for the source ions.
Therefore, using an isotope with a higher 
endpoint energy allows for a lower 
$\gamma$ to obtain the same neutrino energies. 
However, in order to get the same flux, 
the useful isotope decays have to be adjusted 
quadratically. For our pairs of isotopes, 
we have $E_0^{B+Li} \simeq 3.5 \cdot E_0^{Ne+He}$. 
Therefore, from \equ{cond} we obtain the conditions
\begin{equation} 
N_\beta^{B+Li} \simeq 12 \cdot
N_\beta^{Ne+He} \, , \quad
\gamma^{Ne+He} \simeq 3.5 \cdot \gamma^{B+Li}
\label{equ:condnum}
\end{equation} in order
to obtain the same neutrino flux spectrum. 
Note that the number of useful ion decays $N_\beta$ represents, 
to first approximation, 
an {\em ion source degree of freedom}, whereas
the $\gamma$ is an {\em accelerator degree of freedom}.\footnote{There is, 
however, a
non-negligible effect of $N_\beta$ on the 
accelerator by increasing the number of
ions per bunch (or the number of bunches).}
Each can be adjusted with completely different technical challenges.
The actual optimization between higher $\gamma$ versus higher isotope
production rates depends on individual cost and machine aspects, 
and cannot be done at this place \cite{mats_talk_RAL_2008}. 

The optimal baseline depends crucially on the choice of source ions 
and the boost factor. For shorter baselines one is 
away from the matter resonance, and hence 
the flux arriving at the detector is proportional to $1/L^2$. 
If one wants to stay at the oscillation maximum in vacuum, one has
$L/E=const.$, and therefore
$L \propto \gamma$. Since the cross sections 
are proportional to 
$\sim E \propto \gamma$ for deep inelastic scattering (DIS) processes,
one has an overall $1/L^2 \times \gamma \times \gamma^2 = \gamma$
scaling of the event rates in the DIS regime. 
Close to matter resonance, 
the flux at the detector hardly falls as a function of $L$,
which means that longer baselines 
might be preferred. This qualitative discussion of the baseline
dependence does not take into account the non-trivial dependence of the 
oscillation probabilities on the oscillation parameters,
and the intrinsic degeneracies. It is the purpose of this work
to study this dependence.

From the discussion above we see that 
one has to optimize for the Beta-beam flux itself 
by a judicious choice of: 
\begin{itemize}
\item The types of ions and their end-point energy.
\item Lorentz boost factor, $\gamma$.
\item Number of useful ion decays per year, $N_\beta$. 
\end{itemize}
It is clear that every choice 
of ion, $\gamma$, $N_\beta$ and $L$ will give a 
different physics reach for the experiment. The 
choice of $\gamma$ and $N_\beta$ will determine the 
initial Beta-beam flux for a given choice of the source ions.
This is what we would call the ``input'' of the experiment. 
What finally determines the physics reach of the experiment 
is the number of events seen in the detector, and the potential
to resolve correlations and degeneracies. We will call 
this the ``output'' of the experiment. The aim of course 
is to maximize the ``output''. However, 
there are practical limitations 
on stretching the ``input'' possibilities. 
Keeping these in mind, we study the 
comparative sensitivity reach of 
the greenfield Beta-beam set-ups in two ways:
\begin{enumerate}
\item By fixing the input and comparing the output.
\item By fixing the output and comparing the required input.
\end{enumerate}
We will use approach~(1) in section \ref{sec:baseline}, where 
we make a comparison between the sensitivity reach of 
the experiment using either the 
\br and \li combination or the 
\neon and $^6$He combination, as a function of $L$.
Sections~\ref{sec:gamma} and~\ref{sec:lum} are more
in the spirit of approach~(2).

\subsection{The Detector}

We are interested in measuring the golden channel probability 
$P_{e\mu}$. Since we have a $\nue$ ($\anue$) flux in the  
beam, we need a detector which is sensitive to 
muons (antimuons). The detector should have a suitable 
energy threshold, depending on the energy spectrum of the 
Beta-beam. In addition, it should have a good energy 
resolution and low backgrounds. 
There are a number of detector technologies that 
have been considered in the literature. For the 
low energy Beta-beams, water \chr detectors are the 
most widely chosen, mainly because of their low energy threshold and 
large size. This is a very well known and tested detector 
technology. In addition, the detector can be relatively
easily upgraded; typically megaton-sizes~\cite{uno,hk,memp} appear
in the literature. 
However, the backgrounds in this detector 
are generally larger than in other detector types. For 
an intermediate $\gamma$ Beta-beam,
a Totally Active Scintillator Detector (TASD) is a possible
technology. 
This is the option chosen and studied by the 
NO$\nu$A collaboration~\cite{nova}. The third kind of 
detector technology, 
which has been studied extensively and which is 
currently being used by MINOS, is the magnetized iron 
calorimeter. A larger version is envisaged to come up 
soon at the INO facility in India~\cite{ino}. In this paper
it will be referred to as ICAL@INO.
For both TASD and magnetized iron detectors the background 
rejection is typically considered to be better than for water \chr 
detectors. 

%
\begin{table}[t]
\begin{center}
\begin{tabular}{|l|c|}
\hline
&\\[-0.5mm]
Total Mass & 50 kton  \\[2mm]
Energy threshold & 1 GeV \\[2mm]
Detection Efficiency ($\epsilon$) & 80\% \\[2mm]
Charge Identification Efficiency ($f_{ID}$)& 95\%\\[2mm]
Detector Energy Resolution function ($\sigma$)& 0.15E\\[2mm]
Bin Size & 1 GeV\\[2mm]
NC Background Rejection & 0.0001\\[2mm]
Signal error & 2.5\%\\[2mm]
Background error & 5\%\\[2mm]
\hline
\end{tabular}
\caption{\label{tab:detector}
Detector characteristics for neutrinos/antinuetrinos used in the
simulations. The bin size is kept fixed, while the number of bins
is varied according to the maximum energy.
}
\end{center}
\end{table}
%
In this paper we use, for the sake of simplicity, only one 
type of detector for both types of ions and 
all values of $\gamma$. 
We use an ICAL@INO type of detector configuration~\cite{ino}. 
We give the details of our detector specifications in 
Table~\ref{tab:detector}. The charge identification 
efficiency is incorporated since that helps in 
reducing the neutral current backgrounds. 
The number of (anti)muon events 
in the detector is given by 
\be
N_{i} = T\, n_n\, f_{ID}\,\epsilon~  \int_0^{E_{\rm max}} dE
\int_{E_{A_i}^{\rm min}}^{E_{A_i}^{\rm max}}
dE_A \,\phi(E) \,\sigma_\numu(E) \,R(E,E_A)\, P_{e\mu}(E) \, ,
\label{equ:events}
\ee
where $T$ is the total running time (taken as five years),
$\phi(E)$ is the unoscillated flux at the detector, 
$\epsilon$ is the detector efficiency,  
$n_n$ are the number of target nucleons in the detector,
$f_{ID}$ is the charge identification efficiency and  
$R(E,E_A)$ is the energy resolution function of the detector,
for which we assume a Gaussian 
function. For muon (antimuon) events, $\sigma_\numu$ is the 
neutrino (antineutrino) interaction cross-section. 
The quantities $E$ and $E_A$ are the true and 
reconstructed (anti)neutrino 
energy respectively. 

\subsection{Neutrino Propagation and Simulation Details}

The expression for  
$P_{e\mu}$ in matter~\cite{msw1,msw2,msw3}, 
up to second order terms in 
the small quantities $\theta_{13}$ and 
$\alpha \equiv \ms/\ma$, 
is given by~\cite{golden,freund}
\be
 P_{e\mu} &\simeq& 
 \sin^2\theta_{23} \sin^22\theta_{13}
\frac{\sin^2[(1-\hat{A})\Delta]}{(1-\hat{A})^2}\nonumber \\
&\pm& \alpha \sin2\theta_{13} \sin2\theta_{12} \sin2\theta_{23} 
\sin\deltacp \sin(\Delta) \frac{\sin(\hat{A}\Delta)}{\hat{A}}
\frac{\sin[(1-\hat{A})\Delta]}{(1-\hat{A})} \nonumber \\
&+& \alpha \sin2\theta_{13} \sin2\theta_{12} \sin2\theta_{23} 
\cos\deltacp \cos(\Delta) \frac{\sin(\hat{A}\Delta)}{\hat{A}}
\frac{\sin[(1-\hat{A})\Delta]}{(1-\hat{A})} \nonumber \\
&+& \alpha^2 \cos^2\theta_{23} \sin^22\theta_{12} 
\frac{\sin^2(\hat{A}\Delta)}{{\hat{A}}^2}
,
\label{equ:pemu}
\ee
where 
\be
\Delta\equiv \frac{\ma L}{4E},
~~
\hat{A} \equiv \frac{A}{\ma},
\label{equ:matt}
\ee
and $A=\pm 2\sqrt{2}G_FN_eE$ is the matter potential, 
given in terms of the electron density $N_e$ and 
(anti)neutrino energy $E$; the plus sign 
refers to neutrinos while the minus to antineutrinos. 
The second term 
in Eq. (\ref{equ:pemu}) is CP violating. 
While we will use this formula
to discuss our results in some cases, our simulation
is based on the exact probabilities.

\begin{table}
\begin{center}
\begin{tabular}{|l||l|}
\hline
& \\[-.5mm]
$|\Delta m^2_{31}{\rm (true)}| = 2.5 \times 10^{-3} \ {\rm eV}^2$ 
& $\sigma(\Delta m^2_{31})=1.5\%$ 
\\[2mm]
$\sin^2 2 \theta_{23}{\rm (true)}| = 1.0$ 
& $\sigma(\sta)=1\%$ 
\\[2mm]
$\Delta m^2_{21}{\rm (true)} = 8.0 \times 10^{-5} \ {\rm eV}^2$ 
& $\sigma(\Delta m^2_{21})=2\%$ 
\\[2mm]
$\sin^2\theta_{12}{\rm (true)} = 0.31$ 
& $\sigma(\sss)=6\%$ 
\\[2mm]
$\rho{\rm (true)} = 1~{\rm (PREM)}$ & $\sigma(\rho)=5\%$\\[2mm]
\hline
\end{tabular}
\caption{\label{tab:bench}
Chosen benchmark values of oscillation
parameters and their $1\sigma$ estimated errors. The last row
gives the corresponding values for the Earth matter density.}
\end{center}
\end{table}

Unless stated otherwise, we have generated our simulated
data for the benchmark values  in 
the first column of Table~\ref{tab:bench}.
These values have been chosen in conformity with the 
status of the oscillation parameters in the light of 
the current neutrino data~\cite{limits}. The 
values of $\stcht$, $\dcpt$ and mass hierarchy which  
are allowed to vary in our study, will be mentioned
wherever applicable. For the Earth matter density, we use the 
PREM profile~\cite{prem}. 
We expect to have a better 
knowledge of all the parameters mentioned in Table~\ref{tab:bench} 
when the Beta-beam facility comes up. In particular, 
we assume that the $1\sigma$ error on them will be reduced 
to the values shown in the second column of 
Table~\ref{tab:bench}~\cite{huber10,solarprecision,tomography}.
Therefore, 
we impose ``priors'' on these quantities, with the 
corresponding $1\sigma$ error. 
The results presented in section \ref{sec:baseline} 
have been generated using the $\chi^2$ technique 
and numerical code described in 
\cite{betaino1,betaino2}. 
Figures in sections \ref{sec:gamma} and 
\ref{sec:lum} have been generated using the GLoBES 
package~\cite{globes}. 
For the latter simulations we do not put any priors on 
$|\ma|$ and $\stch$ and instead 
add 10 year prospective disappearance data from T2K 
\cite{t2ksimulation}. All other details of 
the $\chi^2$ technique, as well as beam and 
detector specification,  
are taken identical in both numerical codes. 
We have made extensive checks, and 
the results obtained from both codes match to a 
reasonably high level of accuracy. This robustness
of the results can be regarded as an independent
cross-check within our study.

\vglue 1cm
\section{Optimizing the Baseline}
\label{sec:baseline}

We optimize the Beta-beam experiments separately 
with respect to the following physics outputs:
\begin{enumerate}
\item The $\theta_{13}$ measurement reach.
\item The mass hierarchy reach.
\item The CP sensitivity reach.
\end{enumerate}
By ``reach'' we refer to going to as small $\stheta$
as possible. We define the performance indicators below,
and optimize the 
Beta-beam experiment with respect to the 
baseline in this section. 
Note that we fix the $\gamma$ as well as the number of 
useful ion decays to their reference values in 
this section. 

\subsection{The {$\mathbf \theta_{13}$} Sensitivity/Discovery Reach} 
\begin{figure}[t]
\begin{center}
\includegraphics[width=10.0cm]{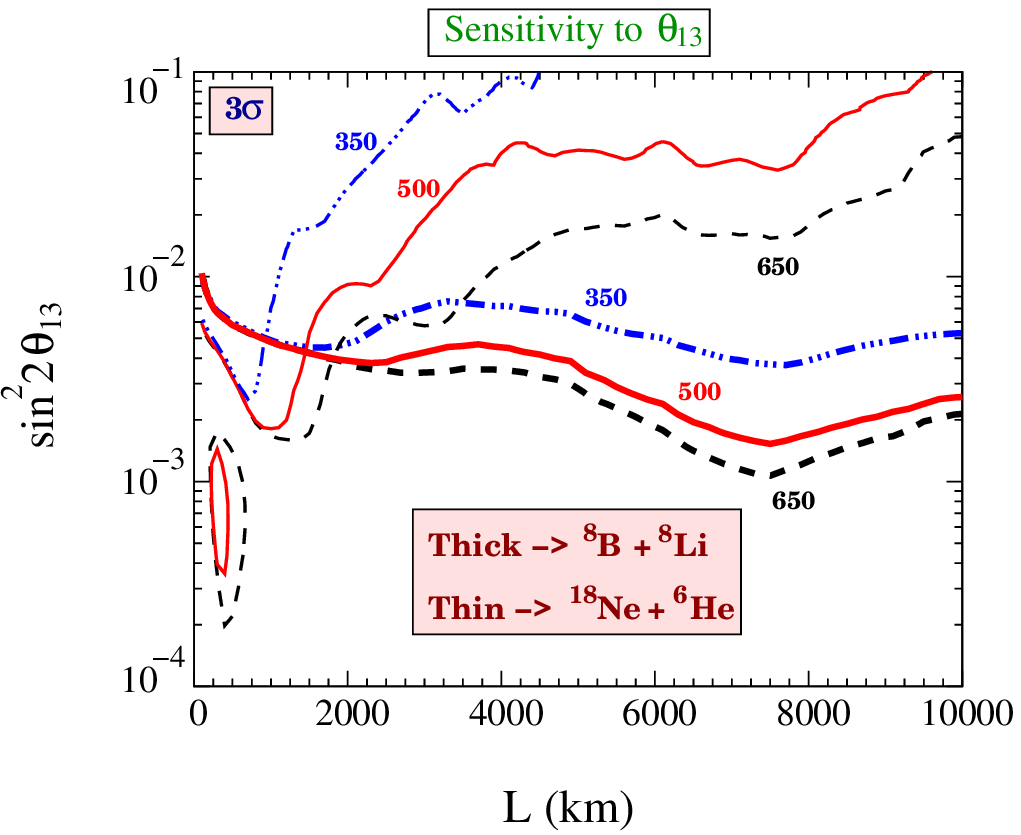}
\caption{\label{fig:sens13L}
The $\theta_{13}$ sensitivity reach 
as a function of the baseline for three different values of
$\gamma=350$ (blue dot-dashed lines), 500 (red solid lines)
and 650 (black dashed lines). 
The thick lines show the results for \br and \li as 
source ions, while the thin lines give the corresponding 
results for \neon and $^6$He. The region above the curves/within
the isolated islands are permitted by the sensitivity criterion.
}
\end{center}
\end{figure}
%
\begin{figure}[t]
\includegraphics[width=0.49\textwidth]{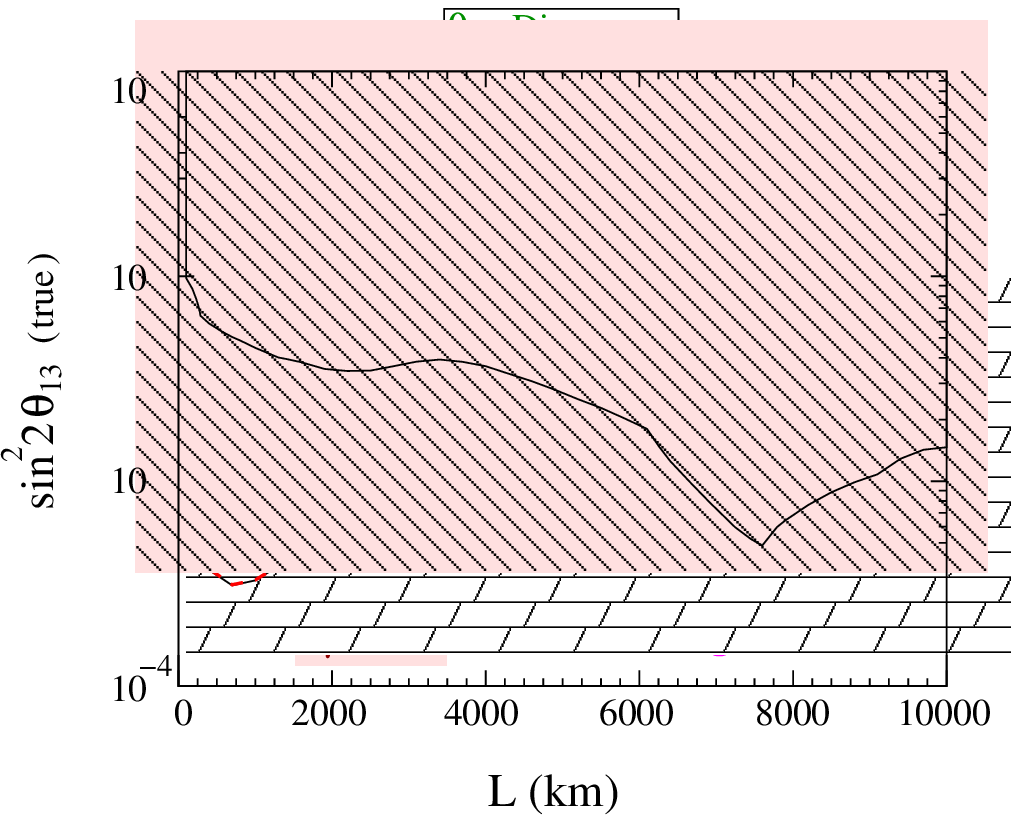}
\includegraphics[width=0.49\textwidth]{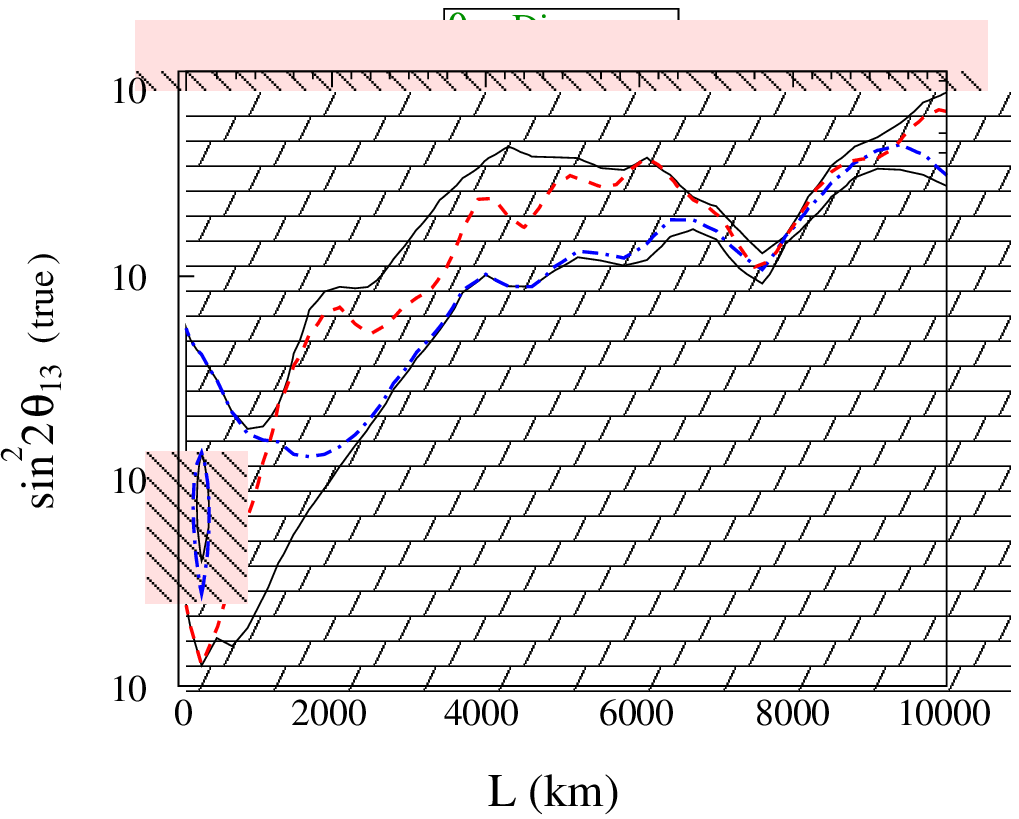}
\caption{\label{fig:discoveryL}
The $\theta_{13}$ discovery reach 
as a function of baseline for \br and \li (left panel) and 
for \neon and \he  (right panel). 
The pink hatched region is where a 
$3\sigma$ discovery is possible for all values 
of $\dcpt$, the white boxed region is where a 
$3\sigma$ discovery is possible for some values of 
$\dcpt$, while the unshaded blank region is where its impossible 
to get a discovery at $3\sigma$ for any value of $\dcpt$.
The red dashed curve is 
the discovery reach for $\dcpt=0$ while
the blue dashed-dotted curve is the same
for $\dcpt=\pi$. Here the true hierarchy is assumed to be normal (NH).
}
\end{figure}

We define two sets of performance indicators for 
quantifying the sensitivity of the experiment to 
$\theta_{13}$. We call them the ``$\theta_{13}$ sensitivity 
reach'' and the ``$\theta_{13}$ discovery reach''. 
The $\theta_{13}$ sensitivity 
reach is defined as the range of $\stch$ 
which is incompatible with the data generated for 
$\stcht=0$ at the $3\sigma$ CL. This performance indicator 
corresponds to the new $\stheta$ limit if the experiment does not 
see a signal for $\theta_{13}$-driven oscillations\footnote{Note 
from \equ{pemu}, while the first three terms go to 
zero when $\theta_{13} \rightarrow 0$, the 
last term, which depends only on the 
solar parameters and $\theta_{23}$, remains non-vanishing. 
Therefore, when the 
flux is high, \ie, for large $\gamma$ and/or enhanced
luminosity, we expect a sizable number of events even when 
$\stcht=0$.}. 
In that case, we can exclude some allowed values of $\stch$, which we call our
 ``$\theta_{13}$ sensitivity reach''.  
In \figu{sens13L}, we show the $L$-dependence of the 
$\theta_{13}$ sensitivity reach of the Beta-beam 
experiment.  The thick lines show the sensitivity 
for the \br and \li combination, while the 
thin lines show the corresponding sensitivities 
for the \neon and \he ions. 
The results are shown for three different values of 
$\gamma$. Since the true value of 
$\theta_{13}$ is assumed to be zero, the data is 
independent of the true neutrino mass hierarchy and $\dcpt$, 
but the fit depends on $\sgnma$ and $\dcp$. We have 
marginalized our results over all oscillation parameters, 
including mass hierarchy and $\dcp$. We have also marginalized over 
the normalization of 
the Earth matter density\footnote{In all results 
given in this paper, we have done full marginalization over 
hierarchy, all oscillation parameters and the 
normalization factor of the Earth matter density distribution.}.
For the \br and \li combination,  
the best $\theta_{13}$ sensitivity is obtained at the 
magic baseline. This baseline is defined by the condition~\cite{magic}
\be
\sin(\hat{A}\Delta) \simeq 0 \, ,
\label{equ:condmagic}
\ee
which evaluates to $\sqrt{2} G_F n_e L(n_e) = 2 \pi$, or
$L \simeq 7 \, 000 $ to $7 \, 500 \, \mathrm{km}$.
Therefore, the second, third and last terms 
in \equ{pemu} vanish at this baseline. Since the second 
and third terms are the CP-dependent terms (with the 
second term being CP violating), the effect of 
$\deltacp$ is absent. Therefore, 
the correlation and degeneracies are hardly present, 
increasing the sensitivity of the 
experiment. 

The impact of the magic baseline is particularly visible 
for the \br and \li combination because for these ions 
the fluxes peak at $E\sim 5-10$ GeV for 
$\gamma \sim 350-650$. It turns out that 
for these energies, one obtains near-resonant 
matter effects, corresponding to $\hat{A} \rightarrow 1$ in \equ{pemu}.
Therefore, the flux decreases less than $1/L^2$ at these
energies, and is still quite substantial at the magic baseline. 
In fact, for short distances, the $1/L^2$ dependence
is canceled by the resonant probability enhancement,
as can be read off \equ{pemu}.
For the \neon and \he combination, 
the fluxes peak around $E\sim 1.0-2.5$ GeV for 
$\gamma \sim 350-650$. For such low energies, matter effects 
are small, even for very long baselines. More importantly,
for $E\sim 1.0-2.5$ GeV, the oscillatory factor peaks at 
$L\sim 500-1250$ km if one assumes 
$\ma=2.5\times 10^{-3}$ eV$^2$. Therefore, for this ion pair,
the minimum in the $\stch$ sensitivity comes at the baseline where 
we expect the first oscillation maximum. For 
$\gamma=650$, 500 and 350, the best sensitivity comes at 
$L=1250$ km, 890 km and 680 km, respectively. Beyond this 
baseline, both the flux and the probability fall, 
resulting in a sharp loss of the $\stch$ 
sensitivity. Note the isolated regions for \neon and \he 
and $\gamma=500$ and $650$, which are also incompatible 
with $\stcht=0$. The gap is mainly 
an artifact of  the presence of clone solutions at 
these smaller baselines and it might
be breached by the combination of a higher flux, better energy
resolution, \etc, leading to a much better sensitivity.

In \figu{discoveryL}, we show the ``discovery reach'' 
for $\stcht$. This performance indicator is defined as the 
range of $\stcht$ values 
which allow us to rule out $\stch=0$ at the $3\sigma$ CL. 
Since the data are now 
generated for a non-zero $\stcht$, there is a $\dcpt$ dependence 
and a true mass hierarchy dependence. The discovery reach for $\dcpt=0$ 
is shown by the dashed curve in \figu{discoveryL},
and the discovery reach for $\dcpt=\pi$ 
is shown by the dashed-dotted curve.
For each $\dcpt$, one obtains a corresponding such curve. 
To show the impact of this $\dcpt$ dependence of the 
discovery potential, and to illustrate explicitly the 
increase in the ``risk factor'' coming from our 
lack of knowledge of $\dcpt$, we present in 
\figu{discoveryL} a band marked by 
boxes showing 
the entire range of $\stcht$ values corresponding to all possible
values of $\dcpt$. \figu{discoveryL} has been drawn for the true
normal hierarchy.

The way to interpret this 
figure is as follows: At a given $L$, one will discover
$\stheta$ for any $\dcpt$ at the upper limit and beyond (the
pink hatched region), whereas there is 
no value of $\dcpt$ for which one can discover $\stheta$ 
below the lower limit (the unshaded region). 
Within the  band, the fraction of $\dcpt$, which allows for a discovery,
increases as one approaches the upper limit. 
Therefore, the upper edge of this band 
gives the most conservative 
discovery reach. We will take 
this as our final discovery reach at a given baseline. 
The lower edge of the band shows the best possible 
case.  The left panel 
of \figu{discoveryL} is computed 
for \br and $^8$Li, while the
right panel is computed for \neon and \he. 
All results in  this figure have been computed for $\gamma=500$. 
Again we note that, at the magic baseline, our results are 
CP independent, and the band reduces to a point 
since the discovery reach is independent of $\dcpt$ 
at this baseline\footnote{For the \neon and \he combination, 
there is a small width even at the magic baseline. In this 
case $\stcht$ is large, and \equ{pemu} 
has to be expanded to higher order. The magic baseline condition
\equ{condmagic} may not hold for higher order terms. This was also 
noted and pointed out in \cite{betaino2}.}. 
For the \br and \li combination 
the 
best discovery reach comes at $L=7600$, which is the magic 
baseline. It is noteworthy though that 
for these ions the best possible 
case, given by the lower edge of the band, 
comes at a lower baseline of $L\simeq 700$ km. 
For the \neon and \he pair, since matter effects are 
low, there is only a local minimum at the magic baseline. 
In fact, the best performance of the experiment 
comes at $L=900$ km, which is approximately the 
baseline where we have the oscillation maximum for 
$\gamma=500$. Again we note the appearance of islands inside 
the band for the \neon and \he ions. 
For regions inside 
these islands, the discovery of $\stheta$ is independent of $\dcpt$. 

We have also computed the $\stcht$ discovery reach for the true inverted
hierarchy. Since the figures look very similar to \figu{discoveryL},
we do not show them explicitely. However, note that in the inverted
case, the curves for $\dcpt=0$ and $\pi$ interchange their roles. It is not
surprising that the inverted hierarchy performs similar to the normal
one for the beta beams, since the neutrino and antineutrino event rates
are very similar. Our chosen number of useful ion decay is about a factor
of three higher for antineutrinos than neutrinos, which is compensated by
the higher neutrino cross sections.

\subsection{The {$\mathbf \sgnma$} Sensitivity Reach} 

\begin{figure}
\includegraphics[width=0.49\textwidth]{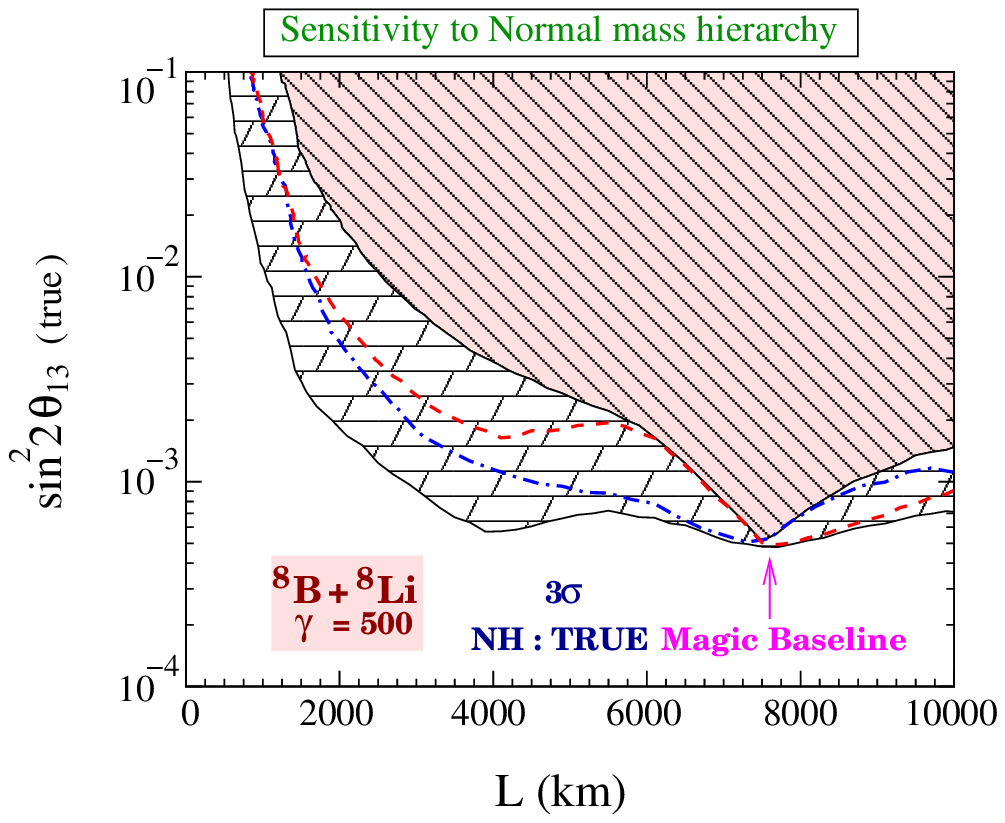}
\includegraphics[width=0.49\textwidth]{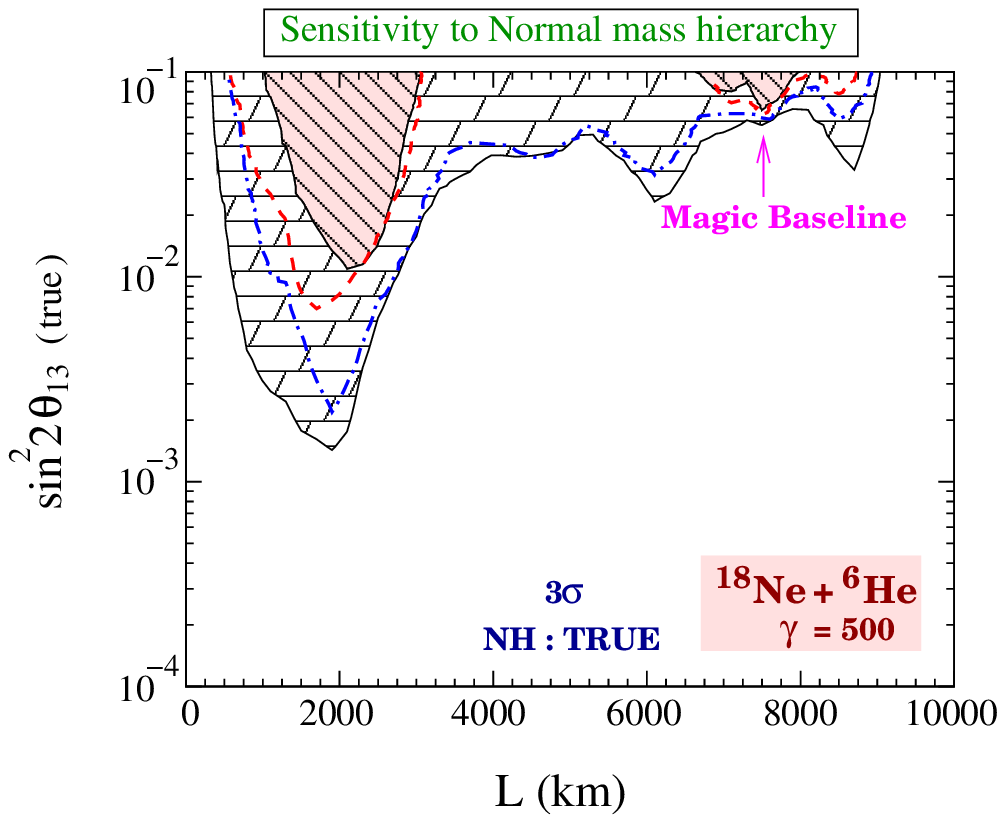}
\caption{\label{fig:hierarchyL}
$\stcht$ reach for determining $\sgnma$ 
at $3\sigma$ CL as a function of the baseline
with \br and \li  (left panel) and
\neon and \he (right panel) taken as the source ions. 
The pink hatched region is where 
the wrong hierarchy can be ruled out 
for all values 
of $\dcpt$, the white boxed region is where the hierarchy 
determination is possible for some values of 
$\dcpt$, while the unshaded blank region is where its impossible 
to determine the hierarchy for any value of $\dcpt$.
The red dashed (blue dashed-dotted) curves  
show the hierarchy sensitivity for $\dcpt=0$ ($\dcpt=\pi$).
The true
hierarchy is assumed to be normal (NH).
}
\end{figure}

We define the mass hierarchy sensitivity as the range  
of $\stcht$ for which the wrong hierarchy can be excluded 
at the $3\sigma$ CL. 
We show our results as a function of the baseline in 
\figu{hierarchyL}. As before, the left panel is for the
\br and \li case, and the right panel for \neon and $^6$He. 
In addition, we show the risk with respect to $\dcpt$ as a band
marked by boxes,
where the lower edge corresponds to the best possible reach (obtainable
for only some specific $\dcpt$),
and the upper edge to the conservative case (valid irrespective
of the value of $\dcpt$). That means
that the hierarchy can be determined for any $\stcht$ above the
upper end of the band. In all panels, we also show the 
curves corresponding to $\dcpt=0$ and $\dcpt=\pi$, for illustration. 
For the \br and \li case, we find that the best sensitivity 
to the mass hierarchy comes at the magic baseline. The 
reason for this is basically the same as in the previous 
subsection: The near-resonant matter effects lead to
a large number of events, and the resonant behavior is only
present for one hierarchy (normal or inverted). 
It is, therefore, possible to have very large 
matter dependent oscillations 
for $L\gtap 4000$ km. However, the effect of 
$\dcpt$ could wash away the sensitivity to $\sgnma$. 
For example, for $L\simeq 4000$ km, we do not obtain a very 
good sensitivity for even the best case. 
If one takes into account 
all possible $\dcpt$ values, the sensitivity becomes deteriorated 
significantly. 
At the magic baseline, the dependence on $\dcpt$ is reduced.
Therefore, this baselines provides the best choice to determine the
mass hierarchy. 

The right panel in \figu{hierarchyL} corresponds to the 
\neon and \he case. The sensitivity to the mass hierarchy 
is rather poor for the values of $\gamma$ we have 
adopted here because of the low energies off the matter resonance
(the energies are about a factor of 3.5 lower than for the \br and \li pair). 
 Since the mass 
hierarchy determination crucially depends on matter effects, 
we have very poor sensitivity for this performance indicator, 
at least for the values of $\gamma$ considered here. We will see 
in the next section that this set of ions could start giving 
comparable sensitivity only when the $\gamma$ is increased 
by a factor of three. For $\gamma=500$, the best sensitivity 
comes at $L\simeq 2000$ km. 

Again, we have tested the true inverted hierarchy case,
and we have not found any significant qualitative or quantitative
differences.

\subsection{The CP Sensitivity Reach} 

\begin{figure}
\includegraphics[width=0.49\textwidth]{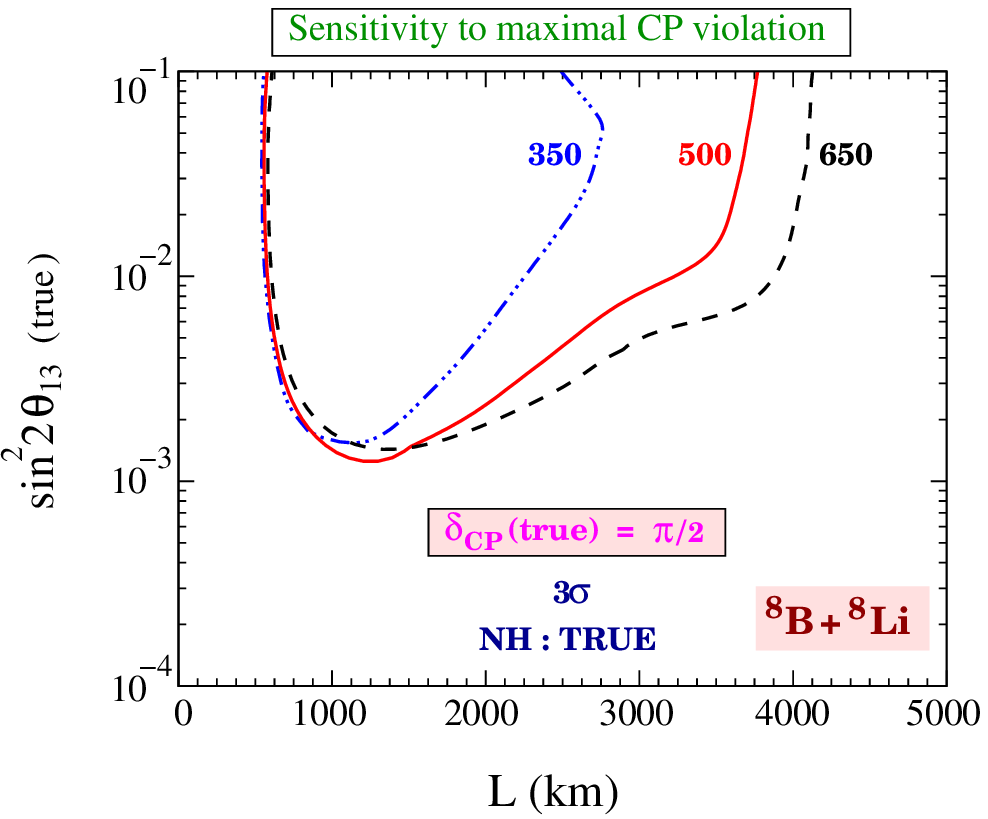}
\includegraphics[width=0.49\textwidth]{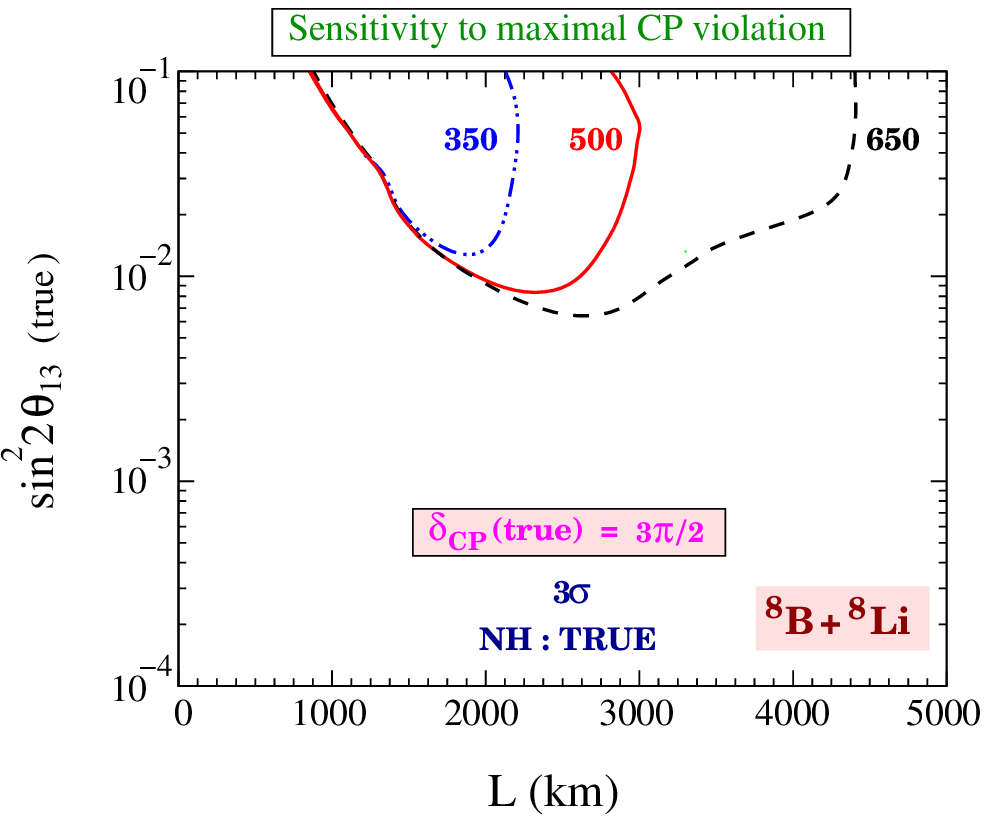}
\includegraphics[width=0.49\textwidth]{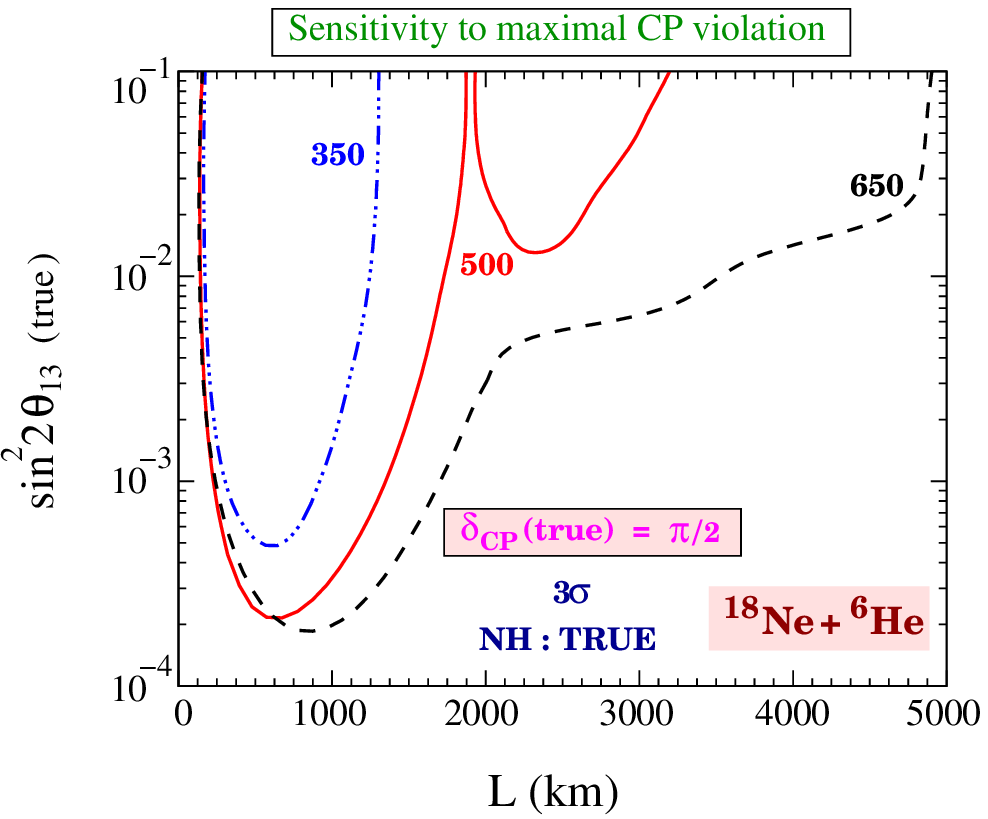}
\includegraphics[width=0.49\textwidth]{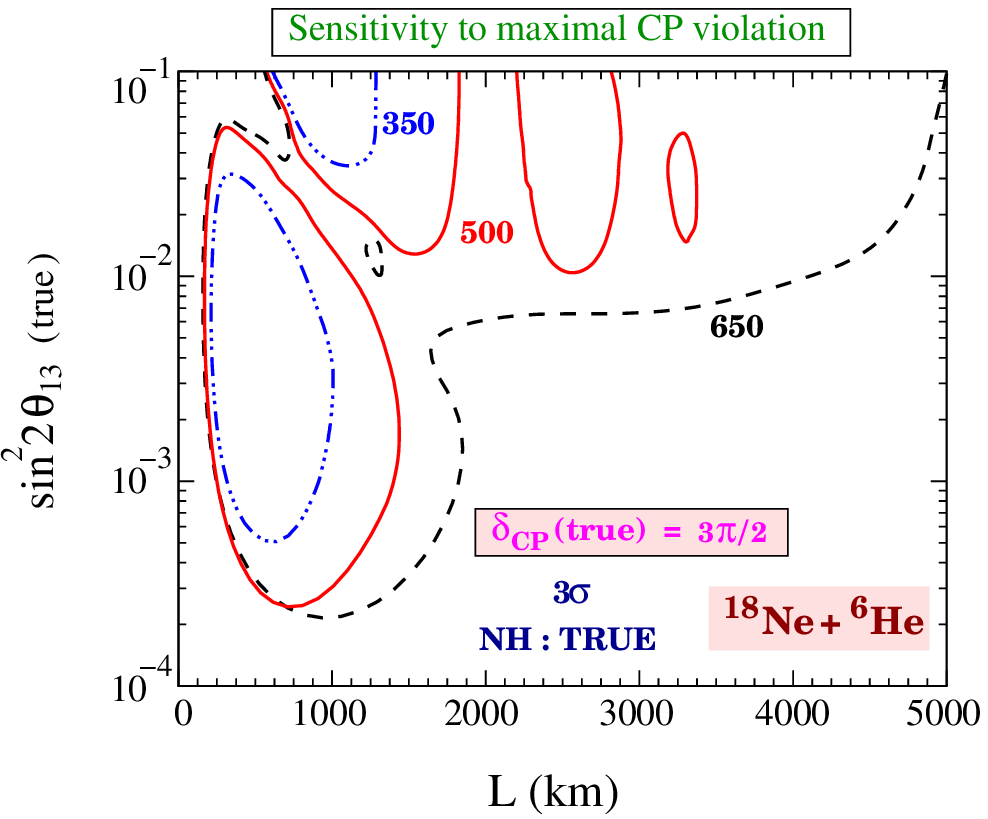}
\caption{\label{fig:cpvL}
$\stcht$ reach for sensitivity to maximal CP violation as a function of $L$, 
for three different values of $\gamma$. 
The upper panels are 
for \br and \li 
taken as the source ions and the lower panels are for 
\neon and \he as the source ions. The left panels are 
for $\dcpt=\pi/2$, and the right panels for 
$\dcpt=3\pi/2$. In all cases, a true normal hierarchy
has been assumed.    
}
\end{figure}

We next discuss the reach of the experiment  
to CP violation as a function of $L$. 
We define the sensitivity to (maximal) CP violation as the
range of $\stheta$ for which CP conservation (\ie, $\dcp=0$ and 
$\pi$) can be excluded at the $3\sigma$ confidence level
irrespective of the (fit) hierarchy.
The results are shown in \figu{cpvL}, where we generate the data 
either for $\dcpt=\pi/2$ (left panels)
or $3\pi/2$ (right panels).
The upper panels  
 are computed for the \br and \li pair, while the lower panels are 
for the \neon and \he pair. 
We show results for three choices of $\gamma$. 
Let us focus on  $\dcpt=\pi/2$ (left panels) first.
The best choice for the 
baseline is around $L=500-1500$ km, depending 
on the choice of ions and $\gamma$.
For example, for  the \br and \li combination,
the optimum is found at around the FNAL-Homestake
baseline $L =1290 \, \mathrm{km}$. 
However, the absolute performance for \br and \li 
is worse than for the \neon and \he combination. 
For $3\pi/2$ (right panels), the sensitivity becomes
worse in both cases due to the impact of degeneracies.
Ignoring the gaps, there is still a 
substantial CP violation reach for \neon and \he  
for both $\dcpt=\pi/2$ and $3\pi/2$. 
For example, if Double Chooz constrains $\stheta$ to values smaller than 
about $0.04$, a short baseline $L \simeq 300$~km together with $\gamma \ge 500$
might be sufficient for the CP violation 
measurement because the gap can be excluded. The optimal baseline 
for maximum $\stheta$ reach, on the other hand, 
ranges between $L\simeq 600$ for $\gamma=350$ to 
$L=1000$ km for $\gamma=650$. For the inverted hierarchy,
we obtain very similar figures to \figu{cpvL}, but the role
of $\dcpt=\pi/2$ and $\dcpt=3\pi/2$ is exchanged.

\subsection{Comparison with a Megaton Water Detector Set-up}

\begin{table}[t]
\begin{center}
{\footnotesize
\begin{tabular}{|c||c|c||c|c||c|c||c|c|} \hline\hline
\multirow{2}{*}{Set-up}
& \multicolumn{2}{|c||}{\rule[0mm]{0mm}{6mm}{$\stch$}}
& \multicolumn{2}{|c||}{\rule[-3mm]{0mm}{6mm}{$\stch$}}
& \multicolumn{2}{|c||}{\rule[-3mm]{0mm}{6mm}{Mass}} 
& \multicolumn{2}{|c|}{\rule[-3mm]{0mm}{6mm}{Maximal}}
\cr
& \multicolumn{2}{|c||}{{\rule[0mm]{0mm}{3mm}Sensitivity}}
& \multicolumn{2}{|c||}{{\rule[0mm]{0mm}{3mm}Discovery}}
& \multicolumn{2}{|c||}{{\rule[0mm]{0mm}{3mm}Ordering}}
& \multicolumn{2}{|c|}{{\rule[0mm]{0mm}{3mm}CP violation, (\sig)}}
\cr
& \multicolumn{2}{|c||}{{\rule[0mm]{0mm}{3mm}(\sig)}}
& \multicolumn{2}{|c||}{{\rule[0mm]{0mm}{3mm}(\sig)}}
& \multicolumn{2}{|c||}{{\rule[0mm]{0mm}{3mm}(\sig)}}
& \multicolumn{2}{|c|}{{\rule[0mm]{0mm}{3mm}$\dcpt=\pi/2$}}
\cr \cline{2-9}
\hline\hline
{\underline{Optimal}}
& \multicolumn{2}{|c||}{{\rule[0mm]{0mm}{6mm}}}
& \multicolumn{2}{|c||}{\rule[-3mm]{0mm}{6mm}{}}
& \multicolumn{2}{|c||}{\rule[-3mm]{0mm}{6mm}{}}
& \multicolumn{2}{|c|}{\rule[-3mm]{0mm}{6mm}{}} 
\cr
ICAL & \multicolumn{2}{|c||}{{\rule[0mm]{0mm}{3mm}$1.5\times 10^{-3}$}}
& \multicolumn{2}{|c||}{{\rule[0mm]{0mm}{3mm}$4.9\times 10^{-4}$}}
& \multicolumn{2}{|c||}{{\rule[0mm]{0mm}{3mm}$5\times 10^{-4}$}}
& \multicolumn{2}{|c|}{{\rule[0mm]{0mm}{3mm}$1.8\times 10^{-4}$}}
\cr
$\gamma = 500$ & \multicolumn{2}{|c||}{{\rule[0mm]{0mm}{3mm}(L=Magic, $^8$B+$^8$Li)}}
& \multicolumn{2}{|c||}{{\rule[0mm]{0mm}{3mm}(L=Magic, $^8$B+$^8$Li)}}
& \multicolumn{2}{|c||}{{\rule[0mm]{0mm}{3mm}(L=Magic, $^8$B+$^8$Li)}}
& \multicolumn{2}{|c|}{{\rule[0mm]{0mm}{3mm}(L=700 km, $^{18}$Ne+$^6$He)}}
\cr
& \multicolumn{2}{|c||}{{\rule[0mm]{0mm}{3mm}}}
& \multicolumn{2}{|c||}{{\rule[0mm]{0mm}{3mm}}}
& \multicolumn{2}{|c||}{{\rule[0mm]{0mm}{3mm}}}
& \multicolumn{2}{|c|}{{\rule[0mm]{0mm}{3mm}}}
\cr
\hline\hline
{\underline{Benchmark}}
& \multicolumn{2}{|c||}{{\rule[0mm]{0mm}{6mm}}}
& \multicolumn{2}{|c||}{\rule[-3mm]{0mm}{6mm}{}}
& \multicolumn{2}{|c||}{\rule[-3mm]{0mm}{6mm}{}}
& \multicolumn{2}{|c|}{\rule[-3mm]{0mm}{6mm}{}}
 \cr
water \chr & \multicolumn{2}{|c||}{{\rule[0mm]{0mm}{3mm}}}
& \multicolumn{2}{|c||}{{\rule[0mm]{0mm}{3mm}}}
& \multicolumn{2}{|c||}{{\rule[0mm]{0mm}{3mm}}}
& \multicolumn{2}{|c|}{{\rule[0mm]{0mm}{3mm}}}
\cr
$\gamma = 350$ & \multicolumn{2}{|c||}{{\rule[0mm]{0mm}{3mm}$5.7\times 10^{-4}$}}
& \multicolumn{2}{|c||}{{\rule[0mm]{0mm}{3mm}$8\times 10^{-4}$}}
& \multicolumn{2}{|c||}{{\rule[0mm]{0mm}{3mm}$1.6\times 10^{-2}$}}
& \multicolumn{2}{|c|}{{\rule[0mm]{0mm}{3mm}$5.2\times 10^{-5}$}}
\cr
L = 730 km & \multicolumn{2}{|c||}{{\rule[0mm]{0mm}{3mm}}}
& \multicolumn{2}{|c||}{{\rule[0mm]{0mm}{3mm}}}
& \multicolumn{2}{|c||}{{\rule[0mm]{0mm}{3mm}}}
& \multicolumn{2}{|c|}{{\rule[0mm]{0mm}{3mm}}}
\cr
$^{18}$Ne+$^6$He & \multicolumn{2}{|c||}{{\rule[0mm]{0mm}{3mm}}}
& \multicolumn{2}{|c||}{{\rule[0mm]{0mm}{3mm}}}
& \multicolumn{2}{|c||}{{\rule[0mm]{0mm}{3mm}}}
& \multicolumn{2}{|c|}{{\rule[0mm]{0mm}{3mm}}}
\cr
\hline\hline

\end{tabular}
}
\caption{\label{tab:comparesetup}
Comparison of the conservative 
sensitivity reaches in $\stch$ for the 
different performance indicators. 
Conservative is defined as the reach in $\stch$ 
irrespective of $\dcpt$ and true mass hierarchy. 
For CP violation we give the reach for 
$\dcpt=\pi/2$. 
The upper row gives the 
sensitivity reaches for the optimal set-ups 
identified in this paper, which use ICAL-type magnetized 
iron detectors. The value of $L$ and the ion source are shown 
in parentheses. Lower row gives the sensitivity for the 
benchmark set-up with water \chr detector. The exposures 
correspond to 50~kton of iron and 500~kton of water with
five years of neutrino and antineutrino runs each.}

\end{center}
\end{table}

In this paper, we have restricted ourselves to magnetized 
iron calorimeters as the far detector technology. 
Detector characteristics such as size, efficiency, 
energy threshold, energy resolution and backgrounds 
affect crucially the choice of the source ions and 
their corresponding boosts. 
Energy threshold is expected to be lower  
for water \chr 
detectors, totally active scintillator detectors, 
and liquid argon TPC. 
As an additional complication, the
reconstructed energies of the 
background events can be different.
Therefore, the 
optimization of the Beta-beam depends on the detector
technology as well. In this paper, we have restricted ourselves to 
a single detector option for all cases, in order
to outline the optimization in terms of boost factor,
baseline, luminosity, and especially the ion pair. 
A comparison of different detector technologies will
follow~\cite{optim2}. 
Here we compare our results to a water \chr detector
set-up for the sake of illustration. We use 
as a benchmark the
set-up from \Ref~\cite{bc}, in which 
 \neon and \he ions with $\gamma=350$, 
$L=730$ km, 
and water \chr detector with 4.4~$\mathrm{Mt \times y}$ statistics
were used as one option.  
Such a detector could be placed at the Gran Sasso laboratory 
in Italy or at Canfranc in Spain. If the source is located 
at CERN the baselines are 730~km and 650~km, respectively. 
In order to compare with the results of 
our optimal set-ups, we have
re-computed the sensitivities 
of this benchmark set-up 
with GLoBES~\cite{globes} 
for the same input assumptions
as in this study and a total luminosity 
corresponding to 500~kton of water with
five years of neutrino and antineutrino runs each\footnote{Especially, 
the different values of $\theta_{12}$ and $\ldm$ compared to 
\Ref~\cite{bc} do have some impact on the sensitivities.
The experiment description is taken from the current GLoBES
distribution, updated with the luminosity numbers for this study.}. 
This set-up returns a $\stheta$ sensitivity 
of $5.7 \times 10^{-4}$, while the $\theta_{13}$ discovery 
reach ranges between $4.8 \times 10^{-5}$ and $8.0 \times 10^{-4}$
depending on $\dcpt$. For two specific choices of $\dcpt=0$ and $\pi$ 
the discovery reaches are $3.1 \times 10^{-4}$ and 
$8.0 \times 10^{-4}$ respectively. Normal mass hierarchy  
discovery reach ranges between 
$2.4 \times 10^{-3}$ and $1.6 \times 10^{-2}$ depending on 
$\dcpt$. For $\dcpt=0$ and $\pi$ 
the hierarchy sensitivity reaches are 
$1.0 \times 10^{-2}$ and $4.0 \times 10^{-3}$ respectively. 
Maximal CP violation can be 
established if $\stcht \geq 5.2 \times 10^{-5}$ for $\dcpt=\pi/2$ 
and $\stcht \geq 5.5 \times 10^{-5}$ for $\dcpt=3\pi/2$.
All numbers at the $3\sigma$ confidence level, and we have not  
found any disconnected regions
for these performance indicators.
For a direct comparison with the optimal set-ups 
identified in this paper, 
we present in Table \ref{tab:comparesetup} the 
``most conservative'' sensitivity reaches in $\stcht$ for 
the four types of performance indicators we have defined. 
``Most Conservative'' is defined as the reach after 
allowing for all possible $\dcpt$ and true mass hierarchy. 
This corresponds to a ``no-risk'' situation. 
The upper row gives the sensitivity reaches for our 
optimized set-ups for the different performance indicators. 
The set-up concerned is defined within the parentheses. 
The lower row gives the corresponding reaches for the 
set-up involving water \chr detector defined above.  
One can see that the water \chr set-up comprehensively 
outperforms our optimal set-up in CP violation, while 
mass hierarchy measurement is clearly done better with \br and 
\li ions at the magic baseline. The $\theta_{13}$ measurement 
is better in water \chr set-up, though our optimal set-up 
is only slightly worse. 
Also, it is unclear
if the 50~kton iron calorimeter is comparable to the 500~kton 
water \chr detector (fiducial volume) in terms of cost.

\section{Optimizing the Baseline and Gamma}
\label{sec:gamma}

In this section, we will allow for larger values of $\gamma$ 
and show results in the $\gamma-L$ plane, while we still 
fix the number of useful ion decays to its reference value. 
In section 2, we have compared the different ion pairs for the same
fixed $\gamma$. This however, implies that the peak energies,
and therefore the physics (matter effects \etc) is different. 
Here we wish to show the projected sensitivity for the 
two sets of isotopes when they produce beams with 
similar peak energies, where we leave both $L$ and $\gamma$
as free parameters. The motivation is to have similar 
effect of the oscillations and matter effects for
both the sets of ions.  The end-point energies of 
\neon and \he are about 3.5 times smaller than those of  
\br and \li. Therefore, with $\gamma$ for \neon and \he 
scaled up by a factor of roughly 3, 
we expect almost the same oscillated spectral shape  
for the two cases. 
Hence we show the results  
for $\gamma < 1000$ for \br and $^8$Li, and for
$\gamma < 3000$ for the \neon and $^6$He.  
We stress that we allow for these prohibitively
large values of $\gamma$ for \neon and \he  
in order to compare set-ups with same physics output. 
While they might appear to be unrealistic, they can 
in principle be achieved by an 
accelerator as large as the LHC \cite{matsmoriond}.
The Tevatron might also be used to give 
large boost factors.  

\subsection{The {$\mathbf \theta_{13}$} Sensitivity} 

\begin{figure}[tp]
\begin{center}
\includegraphics[height=0.65\textheight]{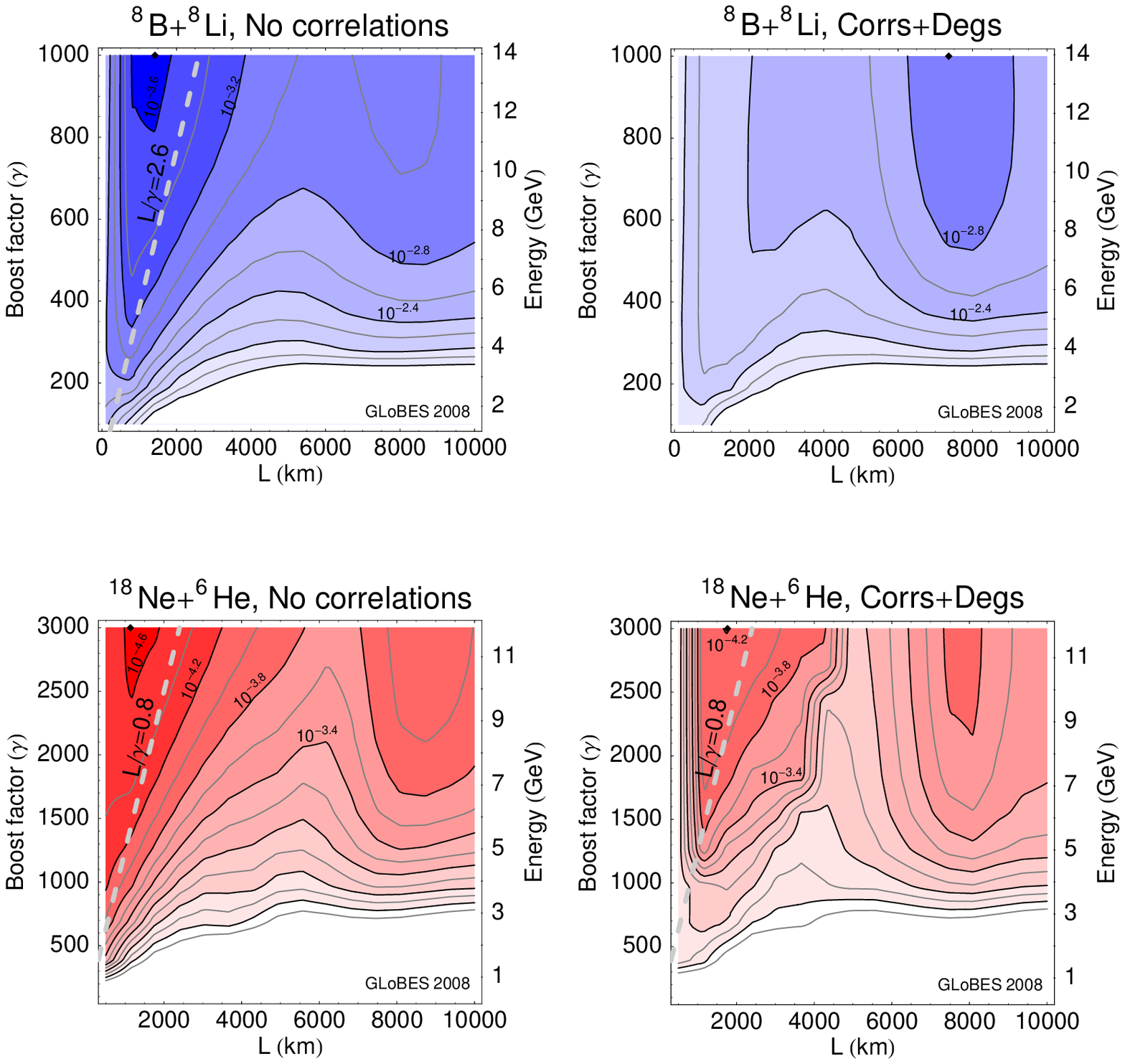}

\vspace*{-1cm}

\end{center}
\caption{\label{fig:isocomp}The $3\sigma$ $\stch$ sensitivity as a
function of baseline $L$ and boost factor $\gamma$.
The $\stch$ sensitivity represents the largest possible 
$\stheta$, which fits data simulated at $\stcht=0$.
The upper row corresponds to $^8$B and $^8$Li, the lower row to
$^{18}$Ne and $^6$He. 
The left column shows the systematics limit only (\ie, the
oscillation parameters are fixed with $\deltacp=0$), whereas the
right column shows the final limit including correlations and
degeneracies. The contours are spaced by $0.2$ in
$\log_{10}(\stheta)$, where the numbers are given for some of
these. The diamonds mark the absolute optimum within each plot,
which are $10^{-3.66}$, $10^{-3.00}$, $10^{-4.66}$, and
$10^{-4.21}$, respectively, from the upper left to the lower
right.  On the right axes of the plots, an energy scale is
attached which corresponds to the mean peak energy $\gamma
\bar{E}_0$ with $\bar{E}_0$ being the mean of the endpoint
energies for the isotope pair.  }
\end{figure}

In order to discuss the $\theta_{13}$ performance, 
we use the $\stch$ sensitivity reach
in this section. 
The $\stheta$ sensitivity represents the {\it largest} fit $\stch$ 
which fits a $\stcht=0$. 
Therefore
in this section, we do not take into account any disjoint regions, 
such as the ones which appeared in \figu{sens13L}. 
We reiterate that the $\stch$ 
sensitivity is independent of the true mass hierarchy 
and $\dcpt$.
We show in \figu{isocomp} the $3\sigma$ 
$\stheta$ sensitivity reach as a function of
baseline $L$ and Lorentz factor $\gamma$. 
The curves in this  figure represent the $3\sigma$ C.L. contours for different 
values of $\stch$.  The contours are spaced 
by $0.2$ in $\log_{10}(\stheta)$, and 
the numbers are shown in the figure for some of them. 
The diamonds mark the absolute optimum within each plot; 
the corresponding values are given in the figure caption. 
The upper row corresponds to $^8$B and $^8$Li and 
the lower row to
$^{18}$Ne and $^6$He. 
The left column shows the systematics limit only, where we have 
kept all parameters including the mass hierarchy and $\dcp=0$ 
fixed in the fit. The 
right column shows the final sensitivity 
limit after including correlations and degeneracies.
From \Sec~\ref{sec:flux}, we have learned that a gamma range for 
$^{18}$Ne and $^6$He 
about three times larger than for $^8$B and $^8$Li
gives similar neutrino peak energies, which is illustrated
by the right vertical axes in the plots. 
This feature is reflected in our choice of the $\gamma$ ranges
in \figu{isocomp}.  
Note that in both the upper and
lower rows of \figu{isocomp} we have used the same 
isotope decay rates, the same
running times, and the same detector simulation. 

Let us first focus on the
systematics limit shown in the left column. 
The two panels for the 
different isotopes look qualitatively very similar, 
but the absolute performance is much better for \neon and \he 
for $\gamma$ values three times larger. 
This is because while we have tuned the oscillation 
probability for both isotopes to be the same, 
the flux increases as $\gamma^2$, and therefore the 
event rate for the \neon and \he combination scales 
up by a factor of about nine for the three times higher
boost factors. 
According to \equ{condnum}, the absolute sensitivities 
for the two isotopes would be more or less identical if 
the $\gamma$ for \neon and \he was turned up by a 
factor of about 3.5 and 
the source luminosity for \br and \li was enhanced
by a factor of about 12. This means that primary
difference between the upper and lower rows in
\figu{isocomp} is the different event rate, but not the
spectral shape and energies.
We can also note 
from the figure that for a 
give set of source ions and for a given $L$ and $N_\beta$, 
the sensitivity of the experiment increases as $\gamma$ 
is increased. 
For the best sensitivity reach, 
the baseline has then to be adjusted accordingly. 
The figure shows that the best choice for 
$L$ roughly corresponds to tuning 
$L/\gamma \simeq 2.6$ for \br and \li and 
$L/\gamma \simeq 0.8$ for \neon and $^6$He.\footnote{In fact,
the unit of $L/\gamma$ is km, which we do not put explicitly.} 
We show the lines for these conditions in
\figu{isocomp} in the upper left panel and lower panels\footnote{Since 
for \br and \li we do not get the optimum at 
the oscillation maxima obeying $L/\gamma \simeq 2.6$ once 
correlations and degeneracies are taken, we do not 
show the line in the upper right
panel.}.  
Note that the slopes of these lines are 
different by about a factor of 
$E_0^{(1)}$/$E_0^{(2)} \simeq 3.5$ because the boost factors 
are related by \equ{condnum} in order to obtain the same neutrino
energies. The optimal regions appear slightly to the left of 
these lines in the figure because statistics are higher at 
lower $L$. 

Once the correlations and degeneracies are included 
(\cf, right column of \figu{isocomp}), 
the optimum baseline changes qualitatively 
for \br and $^8$Li, but not for \neon and $^6$He. 
One can read off from the upper right panel 
that the magic baseline 
becomes the optimum baseline for about $\gamma \ge 350$, 
whereas for $\gamma < 350$ the shorter baseline is preferred. 
The reason for this was also discussed in the previous 
section. For $\gamma \ge 350$ one gets a peak neutrino 
energy greater than 5 GeV. For these energies one can 
obtain very large matter effects for the very long 
baselines and hence the oscillation probability becomes 
larger here. Therefore, the event rate in this regime 
is improved due to a combination of large 
matter driven oscillation probability as well as 
increased flux driven by larger $\gamma$. Most importantly, 
close to the magic baseline, the effect of $\dcp$ is 
almost negligible. Therefore, once the correlations and 
degeneracies are taken into account, this becomes 
the deciding factor which 
ensures that for $\gamma \ge 350$, the magic baseline
emerges as the most optimal baseline choice for 
the $\theta_{13}$ sensitivity.   
For $\gamma < 350$, both the oscillation probability as 
well as the flux collimation are small, and therefore 
the longer baselines suffer due to the $1/L^2$ suppression 
factor for the event rates. As a result, despite being 
free from problems of correlations and degeneracies, the 
magic baseline looses sensitivity compared to the shorter baselines. 
Also note that the sensitivity for $\gamma < 350$ is 
rather poor, and therefore it is desirable to do 
the experiment at $\gamma \ge 350$. 

The situation is very different for the \neon and \he case, 
where the shorter baseline (which roughly satisfies 
$L/\gamma \simeq 0.8$) is always the 
better choice for the $\stheta$
sensitivity for any value of $\gamma$. 
Note that we have adjusted $\gamma$ to keep 
the same peak energy (shown by the dashed lines)
and hence we have the same spectral shape and the same effect 
of the oscillation probability as compared to \br and $^8$Li. 
However, the $\gamma$ here is about three times larger, 
and that becomes the 
overwhelming deciding factor for the most suitable baseline. 
The enhancement in the flux 
due to the beam collimation in this case ensures 
that the shorter baselines have a high enough statistics to handle the 
problem of parameter correlations and degeneracies. 
Therefore, the $\stch$ sensitivity is better than 
that of the magic baseline, which is free from a  
$\dcp$ dependence but suffers from 
the $1/L^2$ suppression. Therefore, the best sensitivity roughly  
follows the $L/\gamma=0.8$ line. 
We remind the reader that the isolated regions obtained for 
$\gamma=500$ and 650 in \figu{sens13L} also correspond to 
$L/\gamma\simeq 0.8$.  
For small values of $\stch$, that are relevant here, 
the $\dcp$ dependent terms in the fit are extremely important 
at values of $L$ close to $L_{oscmax}/2$. 
This corresponds to $L/\gamma\simeq 0.8$ 
for \neon and $^6$He. 

In the following section, we will 
compare the long (magic) with the short baseline. 
For the long (magic) baseline, we will
choose $L=7500 \, \mathrm{km}$ (optimum). 
For the short baseline we will take 
the local optimum obtained, which obeys  
$L/\gamma \simeq 2.6$ for \br and \li and
$L/\gamma \simeq 0.8$ for \neon and $^6$He. 

\subsection{The {$\mathbf \sgnma$} Sensitivity Reach} 

\begin{figure}
\begin{center}
\includegraphics[height=0.65\textheight]{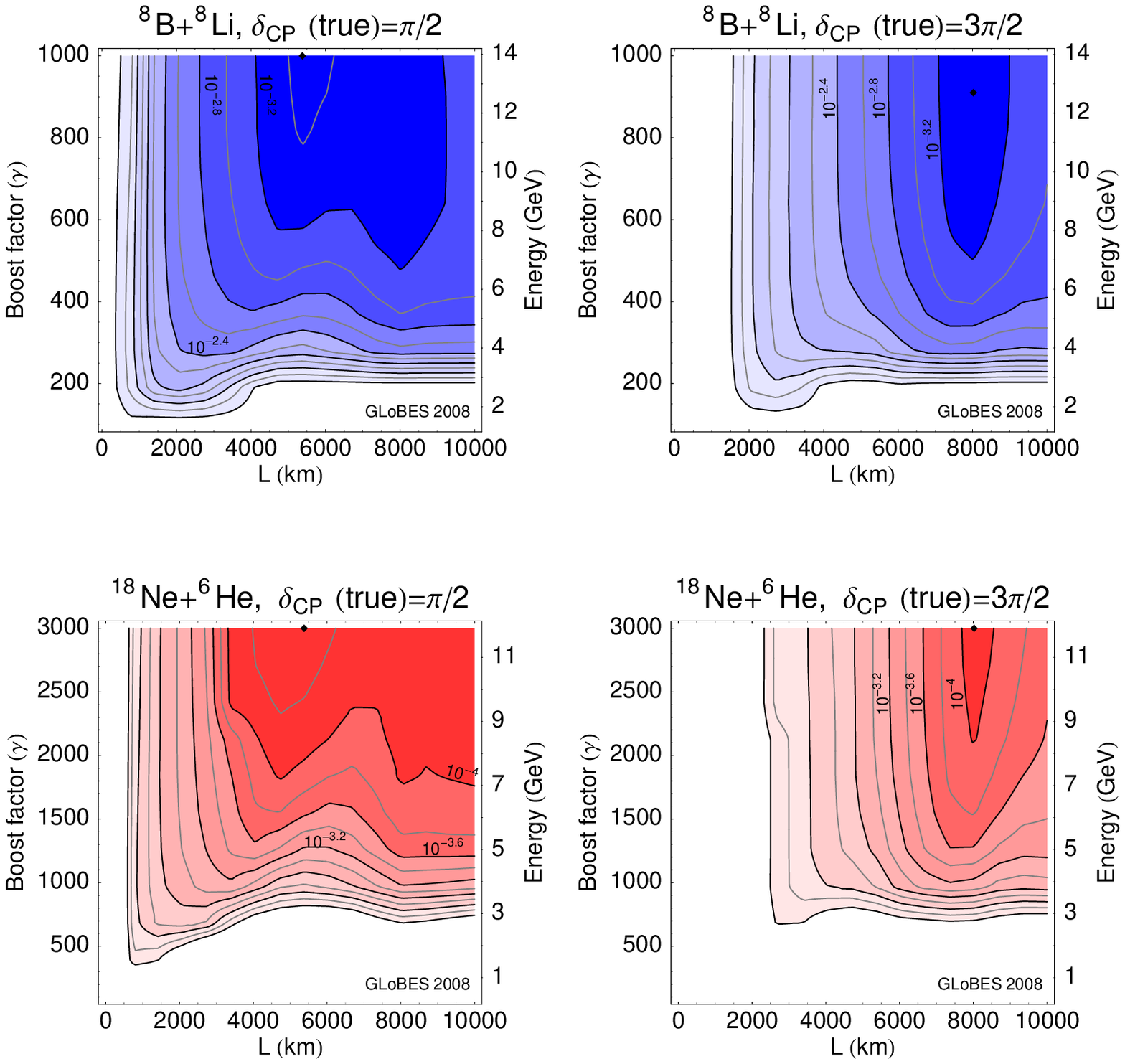}

\vspace*{-1cm}

\end{center}
\caption{\label{fig:isocompmh}The $\stcht$ reach for the sensitivity to the
normal mass hierarchy ($3 \sigma$) as a function of the baseline $L$ and boost 
factor $\gamma$. 
The $\stcht$ reach represents the minimum $\stcht$ above which the mass
hierarchy will be discovered for {\em any} 
$\stcht$ (\ie, there are no gaps
in the sensitivity). 
The upper row corresponds to $^8$B and $^8$Li, while the lower row to
$^{18}$Ne and $^6$He. 
The left column is computed for $\dcpt=\pi/2$, 
whereas the right column is 
for $\dcpt=3 \pi/2$. The
contours are spaced by $0.2$ in $\log_{10}(\stcht)$, 
where the numbers are given for some of these. 
The diamonds mark the absolute
optimum within each plot, which are 
$10^{-3.46}$, $10^{-3.36}$, $10^{-4.29}$, and
$10^{-4.10}$, respectively, from the upper left to the lower right. 
On the right axes of the plots, an energy scale is attached which
corresponds to the mean peak energy $\gamma \bar{E}_0$ with $\bar{E}_0$
being the mean of the endpoint energies for the isotope pair. }
\end{figure}

In \figu{isocompmh}, we show the $\stcht$ reach for the sensitivity 
to the normal mass hierarchy ($3 \sigma$) as a function of baseline 
$L$ and boost factor $\gamma$.
The $\sgnma$ reach represents the minimum $\stcht$ above 
which the mass hierarchy will be discovered for {\em any} 
$\stcht$ (\ie, there are no gaps in the sensitivity).
The upper row corresponds to $^8$B and $^8$Li, the lower row to 
$^{18}$Ne and $^6$He. The left column is computed for  
$\dcpt=\pi/2$, whereas the right column is for 
$\dcpt=3 \pi/2$. 
Note that  the $\sin \delta$ term in \equ{pemu}
is positive for neutrinos and negative for antineutrinos. 
This means that the simulated neutrino rate
is larger for $\dcpt=\pi/2$ and
smaller for $\dcpt=3 \pi/2$. 
Therefore, statistics is better for
$\dcpt=\pi/2$, and we expect a better 
mass hierarchy sensitivity. 
It is for this reason we have chosen to illustrate the 
$\sgnma$ sensitivity reach for 
$\dcpt=\pi/2$ and $3 \pi/2$ only in 
\figu{isocompmh}, which represent two
cases close to the best case and worst case. 
These two cases would change
their role if one used a true inverted 
hierarchy instead, \ie, the best
performance would be close to $\dcpt=3 \pi/2$, and the worst close to
$\dcpt=\pi/2$.

For both isotope pairs, there are two 
main observations for the mass hierarchy
measurement:
\begin{enumerate}
\item
Longer baselines are preferred with the optimal 
sensitivity reach appearing at $L$ 
close to the magic baseline. This is not surprising since for longer
baselines the matter effect contribution becomes larger, which allows to
discriminate between normal and 
inverted hierarchy. The sensitivity gets better at 
the magic baseline since the probability at the 
magic baseline is free of $\dcp$ related correlations and 
degeneracies. 
\item 
Higher boost factors are preferred (at least within the shown ranges). 
First of all, the event rate increases as $\gamma$ increases.
However, matter effects also increase as one approaches the
mantle resonance energy at about 7~GeV. Therefore, 
sufficiently high energies are needed to observe
the mass hierarchy discriminating matter effects.
\end{enumerate}

In summary, we find that the optimal 
choice for determining the hierarchy is 
$L \sim 7000- 9000 \, \mathrm{km}$, 
and $\gamma \gg 200$ for \br and $^8$Li or $\gamma \gg 750$
for \neon and $^6$He. 
Of course, for the same energy, 
the absolute performance is better
for \neon and \he as long as one can create boost factors 
which are three times larger.

\subsection{The CP Sensitivity Reach} 

\begin{figure}[pt]
\begin{center}
\includegraphics[height=0.65\textheight]{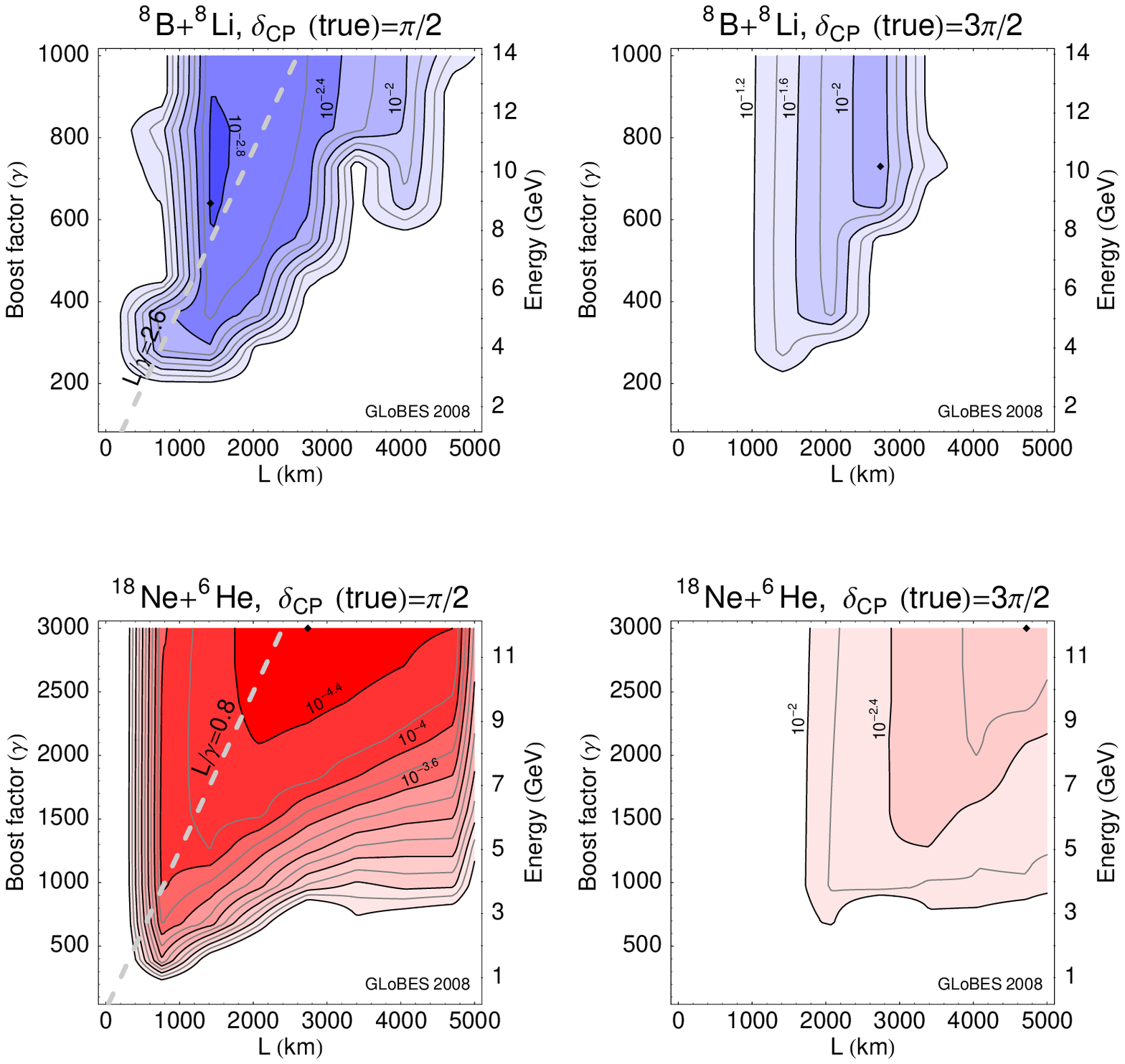}

\vspace*{-1cm}

\end{center}
\caption{\label{fig:isocompcp}
The $\stcht$ reach for the sensitivity to 
maximal CP violation ($3 \sigma$) as a function of 
the baseline $L$ and boost factor $\gamma$.
The $\stch$ reach represents the minimum 
$\stcht$ above which maximal CP violation 
will be discovered for {\em any} 
$\stcht$ (\ie, there are no gaps in the sensitivity). 
The upper row corresponds to $^8$B and $^8$Li, the 
lower row to $^{18}$Ne and $^6$He. The left column 
is computed for $\dcpt=\pi/2$, whereas 
the right column is for $\dcpt=3 \pi/2$. 
The contours are spaced by $0.2$ in $\log_{10}(\stcht)$, 
where the numbers are given for some of these. The 
diamonds mark the absolute optimum within each plot. 
Here a true normal hierarchy is assumed.  The diamonds mark the
absolute optimum within each plot, which are $10^{-2.82}$,
$10^{-2.14}$, $10^{-4.50}$, and $10^{-2.79}$, respectively, from
the upper left to the lower right. On the right axes of the
plots, an energy scale is attached which corresponds to the mean
peak energy $\gamma \bar{E}_0$ with $\bar{E}_0$ being the mean of
the endpoint energies for the isotope pair. }
\end{figure}

In order to discuss the CP sensitivity, we show in \figu{isocompcp}
the $\stcht$ reach for the 
sensitivity to maximal CP violation ($3 \sigma$)
as a function of the baseline $L$ and boost factor $\gamma$. 
The data are generated for normal hierarchy and 
$\dcpt=\pi/2$ (left panels) 
or $3\pi/2$ (right panels).  
We show in the figures 
the minimum $\stcht$ above which maximal
CP violation will be discovered for {\em any} $\stcht$.  
In these figures we discard 
the lower allowed islands in  \figu{cpvL} (lower right panel)
and consider only the upper regions. 
The upper row corresponds to $^8$B and $^8$Li, the lower row to
$^{18}$Ne and $^6$He. 
Obviously, for the CP violation sensitivity, a shorter baseline is a must,
because the magic baseline is not sensitive to $\deltacp$. 
Hence, we only show baselines up to 5000 km 
in this figure, since there is
no sensitivity for longer baselines.
As for the mass 
hierarchy reach, here too the performance for the normal
hierarchy is best for $\dcpt=\pi/2$ and worst for $\dcpt=3\pi/2$.
However, their roles change for the inverted hierarchy.  Note
that the poor sensitivity close to $\dcpt \simeq 3\pi/2$ mainly
comes from unresolved degeneracies due to poor statistics.
Combining data of this set-up with a second (much longer)
baseline could help in resolving the degeneracies and improving
the sensitivity.  Here we focus on the left column of
\figu{isocompcp} for the following discussion.

A comparison of the upper left and lower left panels of
\figu{isocompcp} reveals qualitative differences
between the sensitivities coming from the two pairs of isotopes. 
We observe that the \neon and \he combination 
is far superior for probing the CP phase. We had already seen 
this feature in the earlier section 
when we had compared the CP sensitivity reach of 
the two sets of isotopes for the same $\gamma$. 
In fact, the sensitivity to maximal CP violation improves with 
$\gamma$ for \neon and $^6$He, and we find that the best 
case shown by the diamond appears for the highest $\gamma$ 
we have taken. For \br and \li the sensitivity in general is 
comparatively poorer and does not scale with $\gamma$. In fact, 
the $\gamma$ dependence of the sensitivity is rather weak with 
the best CP sensitivity coming for $\gamma \simeq 650$. 
As far as the possible baselines are concerned, for 
$\dcpt=\pi/2$ the optimal baselines roughly follow the
$L/\gamma=2.6$ line for \br and $^8$Li, 
and the $L/\gamma=0.8$ line for \neon and $^6$He.

\section{The Impact of Luminosity}
\label{sec:lum}

In this section, we study the impact of increasing the 
overall number of events through either increasing the 
size of the detector, the exposure time, detector 
efficiency, or the 
number of useful ion decays. 
In order to discuss this, we introduce a luminosity 
scaling factor multiplying the overall luminosity 
(useful isotope decays $\times$ running time $\times$ detector
mass $\times$ detector 
efficiency) for both neutrinos and antineutrinos. 
Note that the luminosity scaling factor corresponds 
to a reference luminosity, {\it i.e.}, 
$1.1\times 10^{18}$(year$^{-1}$) $\times 5$(year) $\times 50$(kton) 
$\times 0.76$ for the neutrino beam. For the 
antineutrino beam we use  $2.9\times 10^{18}$ useful 
decays per year. 
We attempt to determine the optimal value for 
the luminosity, $\gamma$ and $L$. 
We fix $\gamma$ at certain benchmark values and  
compare the performance of short baselines given by 
$L/\gamma \simeq 2.6$ for \br and \li  
and $L/\gamma \simeq 0.8$ for \neon and $^6$He, 
with the magic baseline. 
For comparison between the isotopes, we use 
both the approaches discussed in section \ref{sec:flux}. 
That is, we compare the physics reach of \br and 
\li with that of \neon and $^6$He, both 
at same fixed values of $\gamma$ corresponding to the 
same input, as well as with $\gamma$ for \neon and \he 
scaled by a factor of about 3.5 to get the same peak neutrino 
energy, and hence the same neutrino energies.
We use the same definition of the performance indicators
as in \Sec~\ref{sec:gamma}, \ie, we demand that there
is sensitivity for all $\stheta$ larger than the given
sensitivity limits (which excludes the regions separated
by the gaps).

\subsection{The {$\mathbf \theta_{13}$} Sensitivity} 

\begin{figure}[tp]
\begin{center}
\includegraphics[width=\textwidth]{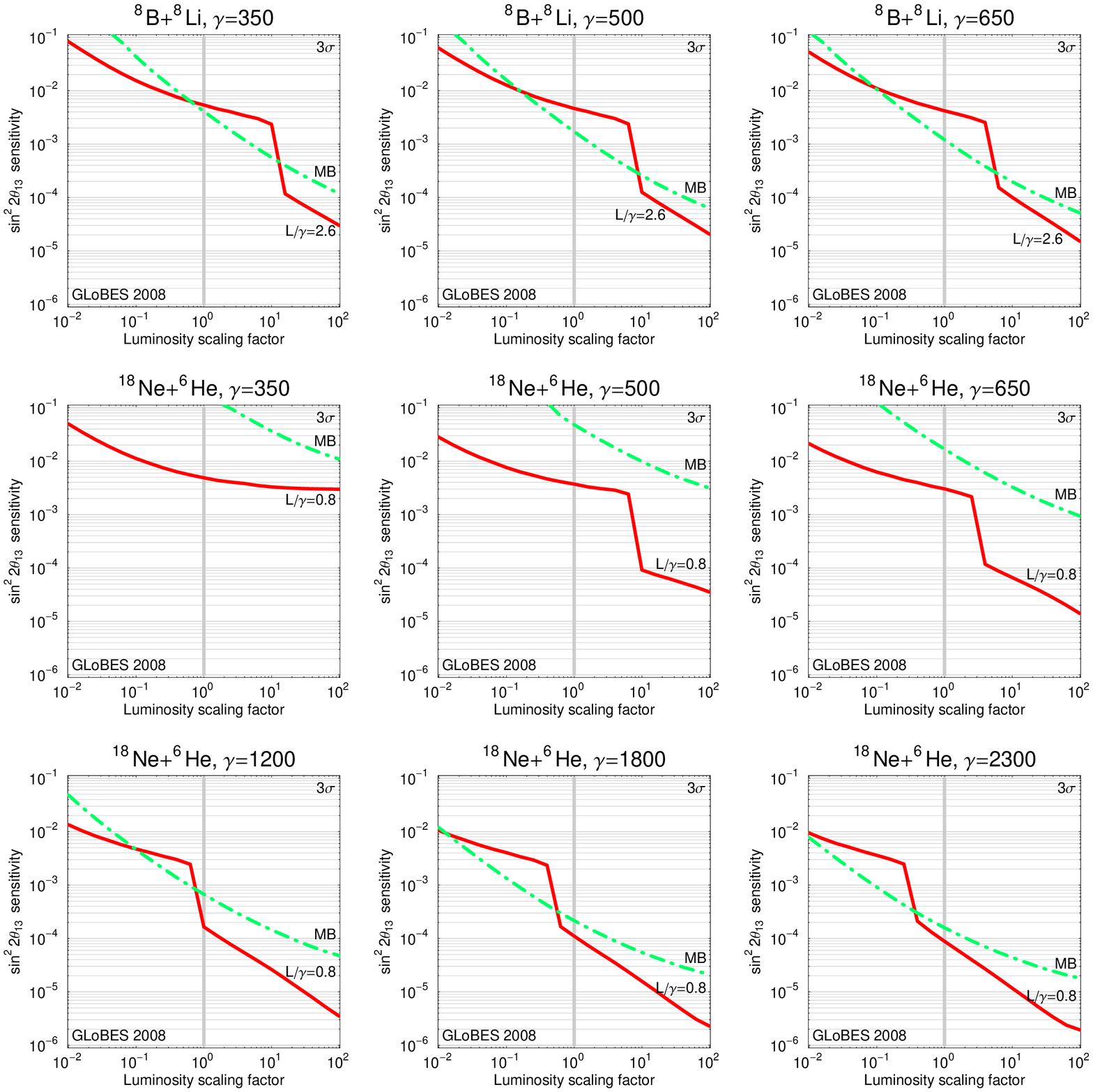}

\vspace*{-0.7cm}

\end{center}
\caption{\label{fig:lumiscale}
The $\stheta$ sensitivity ($3 \sigma$) as a function of a
luminosity scaling factor. The luminosity scaling factor
multiplies the overall luminosities (useful isotope decays
$\times$ running time $\times$ detector mass $\times$ detector
efficiency) for both neutrinos and antineutrinos. The panels
represent the different isotopes and different $\gamma$ as
indicated in the captions. The green dashed-dotted curves
correspond to the magic baseline ``MB'' with $L=7500 \,
\mathrm{km}$ fixed, the red solid curves to a short baseline with
an $L/\gamma$ depending on the isotope (\cf, lines for fixed
$L/\gamma$ in \figu{isocomp}). A true normal hierarchy is assumed.}
\end{figure}

In \figu{lumiscale}, we show the $\stheta$ sensitivity ($3 \sigma$) as a
function of the luminosity scaling factor. The panels represent the
different isotopes and different $\gamma$ values as given in the 
captions. The dot-dashed curves correspond to 
the magic baseline (MB) with
$L=7500 \, \mathrm{km}$ fixed, the solid curves to a short baseline with
an $L/\gamma$ depending on the isotope. The upper row is for 
\br and $^8$Li. The middle and lower rows  
are for \neon and $^6$He, with the middle row for  
same $\gamma$ as \br and $^8$Li, whereas the lower row 
is for $\gamma$'s scaled up by a factor of about 3.5. That means
that the middle row represents the same accelerator effort as the
upper row in terms of $\gamma$, and the lower row  represents
similar neutrino energies to the upper row.

There are a number of interesting 
observations from \figu{lumiscale}. First, 
for the shorter baseline, statistics are 
crucial for resolving the degeneracies, and we obtain a sudden 
enhancement of the sensitivity at some value of the  
luminosity scaling factor. 
This point is visible as the edge in the luminosity scaling,
where the degenerate solution is ruled out at the $3\sigma$ C.L. 
For the magic baseline, degeneracies are hardly  
relevant and we find that the sensitivities exhibit an 
almost power law scaling with statistics. 
Second, the curves for the shorter
baseline cross the ones for the magic baseline twice in almost 
all the panels in the upper and lower rows. For \br and $^8$Li, 
the standard assumed luminosity 
(luminosity scaling factor one) is
typically in the window where the magic baseline 
performs better. Only for
the $\gamma = 350$ case, the shorter baseline is comparable 
with the magic baseline for the standard luminosity.  
For \neon and $^6$He,
the short baseline is typically better for the 
standard luminosity and smaller $\gamma$ (middle row). 
However, for \neon+\he and $\gamma \gg 1000$,
already a factor of two 
loss in luminosity makes the magic baseline the better
choice. 

Let us now come back to the conditions in \equ{cond}. Since we have chosen the
upper and lower rows in \figu{lumiscale} such that the gammas scale inverse to
the endpoint energies, this formula indicated that the same physics should be
obtained for about a factor of~12 difference in luminosity. Let us pick a
simple feature where we could test this conclusion: take a look at the edge of
the sensitivity jumps for the short baselines. In the lower row, this
sensitivity jumps happen for about one order of magnitude less luminosity than
in the upper row, which confirms the expectation. This means that a factor of
12 in the number of useful ion decays is indeed needed to reproduce the same
physics if one uses \br and \li 
instead of \neon and $^6$He. However, this comes for a price: A
factor of 3.5 higher gamma is needed for \neon + \he than for \br
+ $^8$Li.

\subsection{The {$\mathbf \sgnma$} Sensitivity Reach} 

\begin{figure}[tp]
\begin{center}
\includegraphics[width=\textwidth]{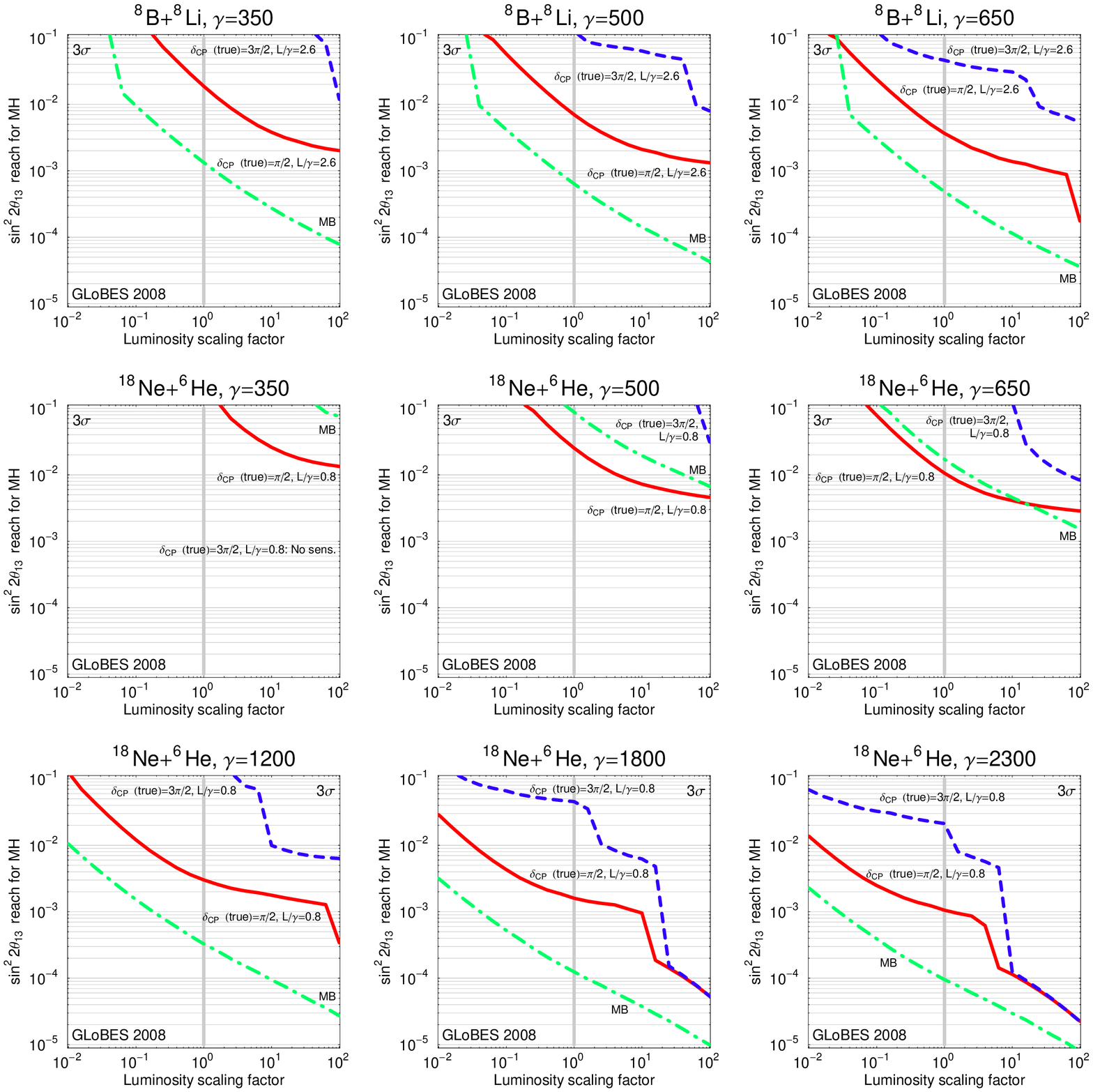}

\vspace*{-1cm}

\end{center}
\caption{\label{fig:lumiscalemh}The $\stheta$ reach for the sensitivity to the
  normal mass hierarchy ($3 \sigma$) as a function of a luminosity scaling
  factor. The luminosity scaling factor multiplies the overall
luminosities (useful isotope decays $\times$ running time
$\times$ detector mass $\times$ detector efficiency) for both
neutrinos and antineutrinos. The panels represent the different
isotopes and different $\gamma$ as indicated in the  captions.
The green dashed-dotted curves correspond to the long magic baseline
``MB'' with $L=7 \, 500 \, \mathrm{km}$ fixed and very little
dependence on the true $\deltacp$, the other two sets of curves
to a short baseline with an $L/\gamma$ depending on the isotope
(\cf, lines for fixed $L/\gamma$ in \figu{isocomp}). The red solid
curves are computed for a true  $\deltacp=\pi/2$, the blue dashed
curves for a true $\deltacp=3\pi/2$.}
\end{figure}

In \figu{lumiscalemh}, we show the $\stheta$ reach for the
sensitivity to the normal mass hierarchy ($3 \sigma$) as a
function of the luminosity scaling factor. The  panels represent
the different isotopes and different $\gamma$ values as given in
the captions. The dashed-dotted curves correspond to the
long magic baseline ``MB'' with $L=7 \, 500 \, \mathrm{km}$ fixed
and very little dependence on the true $\deltacp$, the other two
sets of curves to a short baseline with an $L/\gamma$ depending
on the isotope (\cf, lines for fixed $L/\gamma$ in
\figu{isocomp}). The solid curves are computed for a
$\dcpt=\pi/2$, the dashed curves for a $\dcpt=3\pi/2$.

For the mass hierarchy sensitivity, the magic 
baseline exhibits almost a power law scaling with 
statistics, 
with the best absolute performance in
all cases for all luminosities in the upper and lower rows. For the shorter 
baselines, the performance
depends crucially on the true $\deltacp$. For $\deltacp \simeq \pi/2$, the
scaling behaves similar to the magic baseline, and there are only very few
jumps. That means that changes in the chosen standard luminosity (scaling
factor one) do not have a strong effect. 
Only for the \neon and \he combination with very high $\gamma$,  
increasing the luminosity resolves the 
(wrong hierarchy) intrinsic degeneracy and this 
improves the performance of the 
set-up such that the sensitivities 
approach the ones obtained for the magic baseline. 
For $\deltacp \simeq 3 \pi/2$, the mass hierarchy sensitivity is
basically not present for 
\br and $^8$Li, and for \neon and \he 
for $\gamma \ll 2000$. 
In summary, for the mass hierarchy sensitivity, 
the magic baseline is a safe
choice, independent of the luminosity. Similar 
sensitivity can be
achieved by the shorter baseline only for extreme choices 
of $\gamma$ and $N_\beta$.

\subsection{The CP Sensitivity Reach} 

\begin{figure}[tp]
\begin{center}
\includegraphics[width=\textwidth]{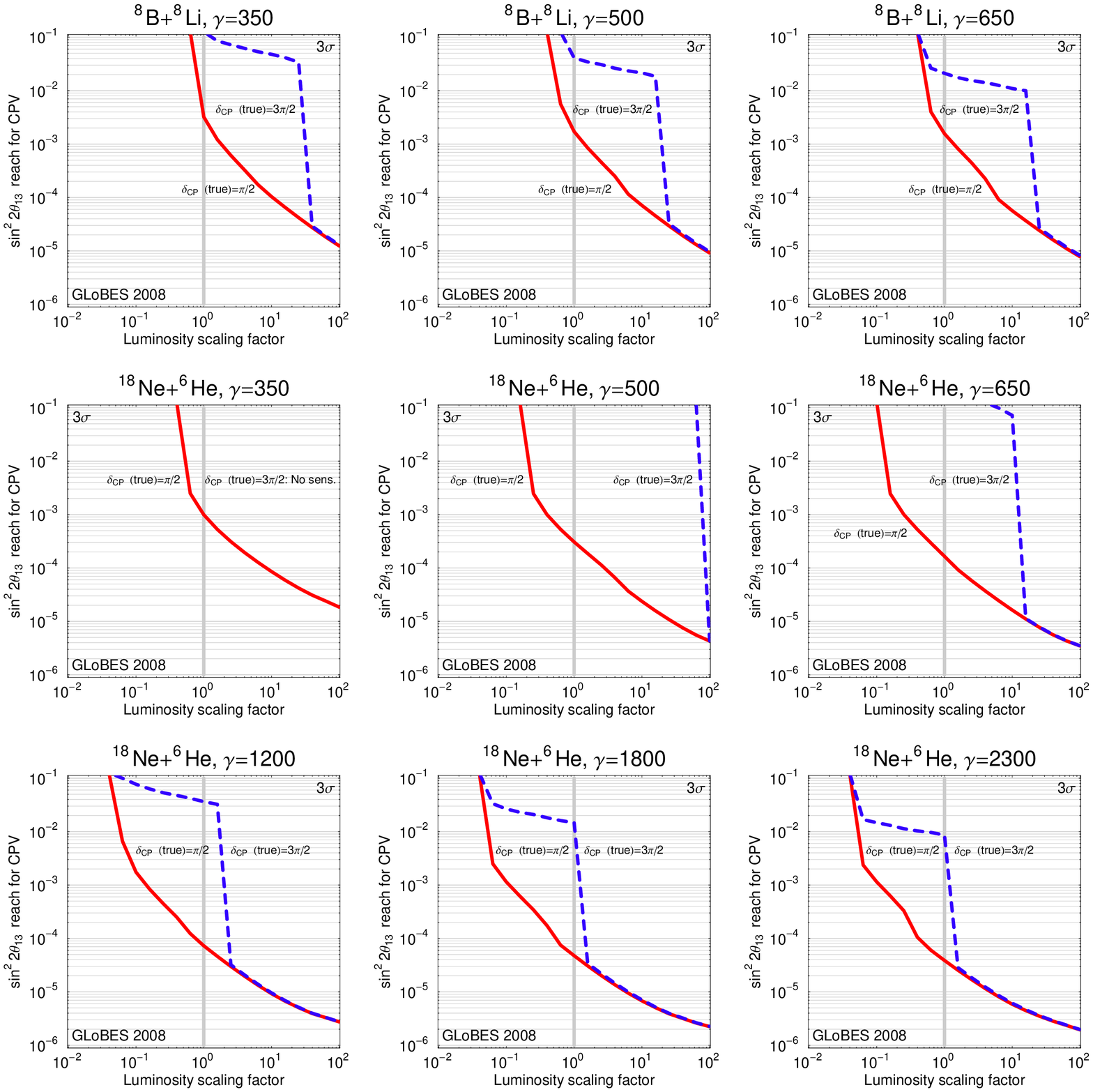}

\vspace*{-1cm}

\end{center}
\caption{\label{fig:lumiscalecp}The $\stheta$ reach for the
sensitivity to maximal CP violation ($3 \sigma$) as a function of
a luminosity scaling factor. The luminosity scaling factor
multiplies the overall luminosities (useful isotope decays
$\times$ running time $\times$ detector mass $\times$ detector
efficiency) for both neutrinos and antineutrinos. The panels
represent the different isotopes and different $\gamma$ as given
in the captions. For all curves, the short baselines are used,
\ie, $L/\gamma=2.6$ for the upper row, and $L/\gamma=0.8$ for the
middle and lower rows.  The red solid curves are computed for a
true  $\deltacp=\pi/2$, the blue dashed curves for a true
$\deltacp=3\pi/2$.  A true normal hierarchy is assumed.}
\end{figure}

The $\stheta$ reach for the sensitivity to maximal CP violation ($3 \sigma$)
as a function of the luminosity scaling factor is shown in
\figu{lumiscalecp}. The luminosity scaling factor multiplies  the overall
luminosities (useful isotope decays $\times$ running time $\times$ detector
mass $\times$ detector efficiency) 
for both neutrinos and antineutrinos. The panels represent the
different isotopes and different $\gamma$ values as given in the 
captions. For all curves, the short baselines are used, \ie, $L/\gamma=2.6$
for the upper row, and $L/\gamma=0.8$ for the middle and 
lower rows, because there is no
CP violation sensitivity at the magic baseline.
The solid curves are computed for $\dcpt=\pi/2$, the dashed curves
for $\dcpt=3\pi/2$.

The scalings are very similar to the mass hierarchy sensitivity: For $\deltacp
\simeq \pi/2$ there are no jumps, and for $\deltacp \simeq 3\pi/2$ the
intrinsic degeneracy can be resolved above a certain threshold luminosity
(because it is then lifted over the $\chi^2=9$ line). All curves look
qualitatively very similar, no matter what $\gamma$ or isotope is
used. However,
The absolute performance is better for \neon + \he 
even for the same $\gamma$. 

As far as the chosen reference luminosity 
(scaling factor one) is concerned, the
\br and \li beam is operated in a region where 
substantial luminosity changes will not
affect the result. However, for \neon and $^6$He, 
already a factor of two to
three luminosity increase would be sufficient to resolve the intrinsic
degeneracy and to boost the performance at 
$\dcpt=3 \pi/2$ for the lower row. Note that such
a boost could also be achieved by a synergistic degeneracy resolver, such as
the magic baseline as second baseline.

\subsection{Comparison with a Neutrino Factory}

As far as the comparison to a Neutrino Factory is concerned,
our Figs.~6 ($\theta_{13}$), 7 ($\mathrm{sgn}(\ma)$), and 8 
(maximal CP violation) 
correspond to Figs.  
3 ($\theta_{13}$), 6 ($\mathrm{sgn}(\ma)$), and 5 (maximal CP violation)
in \Ref~\cite{nufactoptim} for a Neutrino Factory, which means that 
they allow for a direct qualitative comparison.
While for the beta beams, the boost factor is 
on the vertical axis, for a Neutrino Factory, the
muon energy is on the vertical axis. 
Both are proportional to the peak neutrino energy. Comparing
the beta beam and Neutrino Factory qualitatively, 
we find a similar behavior as a function of baseline
and neutrino energy for both isotope pairs and all 
performance indicators, except from \neon and \he 
and the $\theta_{13}$ sensitivity. In this case, the 
shorter baseline dominates, which corresponds to
the correlation only case in Fig.~3 of \cite{nufactoptim}.
Quantitatively, the best performance for the $\theta_{13}$ 
and hierarchy measurements with the Neutrino 
Factory comes at the magic baseline. 
The Neutrino Factory chosen here corresponds to 
$10^{21}$ useful muon decays per year with muon 
energy of 50 GeV and a 50 kton magnetized iron detector.
If the detector efficiency is taken as about 40\% and background 
rejection factor as $1\times 10^{-5}$, 
the sensitivity reaches in $\stcht$ for the two performance 
indicators $\theta_{13}$ and $sgn(\ma)$ at the magic baseline 
are about $2 \times 10^{-4}$ and $1\times 10^{-4}$ 
respectively~\cite{betaoptim}. 
Best CP violation sensitivity comes at $L\simeq 3000-5000$ km. 
At this baseline, the $\stch$ 
reach for maximal CP violation is $7 \times 10^{-5}$~\cite{betaoptim}.
These numbers can be compared with the corresponding reach 
for the optimal Beta-beam set-ups identified in this paper. 
One can see that the Neutrino Factory outperforms 
the Beta-beam for the standard luminosity chosen here and 
for values of $\gamma < 650$. 
However, for higher values of $\gamma$ and higher 
luminosities, Beta-beam returns a performance 
which is comparable to that of a Neutrino Factory. 
Recall that we have taken conservative estimates for 
the number of useful ion decays per year and the 
size of the detector. In principle, it should 
be possible to have at least 10 times more useful decays 
per year \cite{mats_talk_RAL_2008}. In addition, 
for some ranges of $\gamma$ it should be possible to 
use megaton water detectors with Beta-beams, and this 
would again increase the luminosity by factor of 10.

\section{Summary and Conclusions}

Beta-beams provide intense and well understood neutrino fluxes 
of a single flavor. They have very low beam-related 
backgrounds and very low systematic uncertainties,
and are hence ideal for precision 
experiments. While detailed physics reach 
analyses of specific projects involving Beta-beams have been performed
in the literature, greenfield set-ups have been less extensively
studied. In this study, we have considered the 
Beta-beam option as a greenfield 
scenario and have identified the optimal set-ups for an
iron calorimeter as the detector. We have optimized for maximal reaches
in $\stheta$, \ie, our Beta-beams are designed to measure $\stheta$,
CP violation, and the mass hierarchy for as small $\stheta$ as possible.
We have studied 
two pairs of plausible source ions for the Beta-beam: the standard 
ions \neon and $^6$He, which have been extensively studied in the literature,
as well as the new candidates \br and $^8$Li. For each pair of 
source ions, we have optimized the experimental set-up as a function of 
baseline $L$, boost factor $\gamma$, and luminosity. The luminosity 
is proportional to the number of useful ion decays, detector efficiency, 
size of detector, and exposure time. 
We have followed two 
complementary approaches for our study: (i) Fixing 
the input parameters of the beam and looking for the 
sensitivity reach between the two pairs of ions 
as a function of the baseline. (ii) Matching the 
shape (energies) of the event spectrum and luminosity 
for the two 
sets of ions and studying the required input 
parameters of the beam as a function of the baseline. 
In both approaches, we have compared the physics reach for the 
two pairs of ions in order to identify the optimal conditions for 
the sensitivity reach. 

In order to compare the two different pairs of isotopes, it has been useful
to determine the conditions for the same physics output, \ie, neutrino
energies, matter effects, \etc, at the same baseline.
Since the  total flux at the detector for a fixed baseline 
is proportional to $N_\beta \gamma^2$ 
(with $N_\beta$ being the useful ion decays per year) and the peak 
neutrino energy is approximately given by $\gamma E_0$, we   
have identified the conditions for obtaining similar neutrino spectral 
shape as $N_\beta^{B+Li} \simeq 12 \cdot N_\beta^{Ne+He}$, 
$\gamma^{Ne+He} \simeq 3.5 \cdot \gamma^{B+Li}$. This means that,
because of the stronger forward collimation of the beam, 
a smaller required $\gamma$ for \br and $^8$Li has, 
in principle, to be compensated by a 
correspondingly larger number of useful ion decays. We 
have verified this behavior in realistic simulations.

As the next step, we have fixed the beam input parameters $N_\beta$
to $1.1 \times 10^{18}$ ($\nu_e$) and $2.9\times 10^{18}$ 
($\bar{\nu}_e$) useful ion decays/year, and $\gamma$ 
to $350$, $500$, and $650$, respectively.
Because of the higher neutrino energies for the same $\gamma$,
$^8$B and $^8$Li experience stronger matter effects than 
$^6$He and $^{18}$Ne. This has two major implications:
First, if the matter resonance energy can be covered, 
the $1/L^2$ dependence of the events is reduced for $^8$B and $^8$Li,
and the event rates increase at longer baselines. 
Second, because of the matter resonant or anti-resonant behavior,
the mass hierarchy can be much easily determined. Both
 implications make the magic baseline attractive 
for the $\stheta$ and mass hierarchy measurements using 
$^8$B and $^8$Li, has had been shown in 
\cite{paper1,betaino1,betaino2}.
For $^6$He and $^{18}$Ne, on the other hand, a much shorter baseline
is preferable. 
For CP violation however, the $^6$He and $^{18}$Ne turn out to
have a discovery reach \cite{bc,bc2}
about one order of magnitude better than for 
$^8$B and $^8$Li at a relatively short baseline $L \simeq 600-1000$~km
(depending on the $\gamma$ chosen). 
Similar to the Neutrino Factory,
we therefore observe a tension between measuring the mass hierarchy
(preferring long baselines and high energies 
close to the matter
resonance energy), and measuring CP violation (preferring short baselines
determined by the oscillation maximum with as 
little matter effects as possible).
Therefore, in order to optimally access all performance indicators,
one may finally require both pairs of ions \cite{doninialter} and 
two baselines \cite{newdonini}.

If one wants to compare similar neutrino energies for the two isotope pairs,
\ie, one fixes the output, one has to include $\gamma$'s about a factor of
three higher for $^6$He and $^{18}$Ne than for $^8$B and $^8$Li.
From an optimization of our performance indicators in the $L$-$\gamma$ plane
(while still keeping the useful ion decays fixed),
we have learned that there are, in principle, 
two sets of baselines which exhibit
local optima in the performances:
\begin{itemize}
\item
 A short baseline with $L \, 
[\mathrm{km}]/\gamma\simeq 0.8$ for $^6$He and $^{18}$Ne,
and $L \, [\mathrm{km}]/\gamma\simeq 2.6$ for $^8$B and $^8$Li.
\item
 A long ``magic'' baseline 
$L \simeq 7 \, 500$~km, where the dependence on
$\deltacp$ vanishes.
\end{itemize}
While the shorter baseline is always the better choice for the CP violation
measurement, the choice between the longer and shorter 
baseline for the $\stheta$
sensitivity depends on $\gamma$ and the isotope used. For $^8$B and $^8$Li,
the magic baseline is preferred for $\gamma>350$, while 
for  $^6$He and $^{18}$Ne,
the shorter baseline is always preferred for any realistic $\gamma$ because of
the higher event rates for the same neutrino energies.
For the mass hierarchy sensitivity, the magic 
baseline tends to be the best choice
for $^8$B and $^8$Li already for $\gamma>300$, 
while for $^6$He and $^{18}$Ne a relatively
high $\gamma>1000$ is required. However, 
in the latter case, the short baseline alone will 
not be sufficient to measure the mass hierarchy, another longer 
baseline is required.

As the last step, we have focused on the above two sets of baselines,
and we have varied the luminosity. For the $\stheta$ sensitivity,
we have demonstrated that the choice between the short and long baseline
depends on the reference luminosity as well. For the mass hierarchy
sensitivity, the conclusion to use a longer baseline is very robust.
And for the CP violation sensitivity, which is only present at the
short baseline, the luminosity plays a crucial role to resolve
(otherwise present) degeneracies for $\deltacp=3 \pi/2$ (for the
normal hierarchy) or $\deltacp=\pi/2$ (for the
inverted hierarchy). Note that this degeneracy resolution 
could also be achieved
with the combination of two baselines.

We conclude that a greenfield Beta-beam could have excellent
sensitivity reaches for the $\stheta$, mass hierarchy, and CP
violation discoveries. Comparison of the
physics reach between ($^8$B,
$^8$Li) and ($^6$He, $^{18}$Ne) pairs is not
at all straightforward. On the one hand, \br and \li produce
a given neutrino energy by a boost factor about 3.5 times
lower than that needed for \neon and $^6$He. Therefore,
the first pair of ions would be
the preferred choice if one needed higher
energy Beta-beams within the constraint of the envisaged accelerator
facilities.
On the other hand, lower boost factors result in a lower beam
collimation, and hence lower statistics, which would have to
be compensated by a higher luminosity. For
\neon and \he one would need a $\gamma$ about 3.5 times larger than
 for \br and \li to obtain the same neutrino energies. This constraint
stretches
the demanded $\gamma$ to the prohibitively large regime. The statistics,
in return,
would be about a factor of 10 higher.
Therefore, the optimal selection of ions and baselines
crucially depends on the
boost factor and luminosity used,
as well as the chosen detector technology.
For our
reference luminosity and iron detector, if the Beta-beam is
operated at a realistically
``high'' $\gamma \sim 500$, one would optimally use \neon and \he
at the short baseline for CP violation, $^8$B and $^8$Li
at the magic baseline for mass hierarchy, and either $^{18}$Ne
and  $^6$He at the short baseline or $^8$B and $^8$Li at the
magic baseline for $\stheta$ discovery.


\vglue 0.8cm
\noindent
{\Large{\bf Acknowledgments}}\vglue 0.3cm
\noindent
The authors acknowledge the HRI cluster facilities for computation.
This work has been supported by the Neutrino Project
under the XI Plan of Harish-Chandra Research Institute.
W.W. would like to acknowledge support from
Emmy Noether program of Deutsche Forschungsgemeinschaft,
and through DFG grant 446 IND 111/9/07.
In addition, he
would like to thank the members of HRI for their warm
hospitality during his stay.


\begin{thebibliography}{99}

\bibitem{solar}
B.~T.~Cleveland {\it et al.},
Astrophys.\ J.\  {\bf 496}, 505 (1998);
%
J.~N.~Abdurashitov {\it et al.}  [SAGE Collaboration],
J.\ Exp.\ Theor.\ Phys.\  {\bf 95}, 181 (2002)
[Zh.\ Eksp.\ Teor.\ Fiz.\  {\bf 122}, 211 (2002)];
%
W.~Hampel {\it et al.}  [GALLEX Collaboration],
Phys.\ Lett.\ B {\bf 447}, 127 (1999); 
%
S.~Fukuda {\it et al.}  [Super-Kamiokande Collaboration],
Phys.\ Lett.\ B {\bf 539}, 179 (2002);
%
B.~Aharmim {\it et al.}  [SNO Collaboration],
Phys.\ Rev.\ C {\bf 72}, 055502 (2005);
%
C.~Arpesella {\it et al.}
  [Borexino~Collaboration],
  Phys.\ Lett.\  B {\bf 658}, 101 (2008).


\bibitem{kl}
T.~Araki {\it et al.}  [KamLAND Collaboration],
Phys.\ Rev.\ Lett.\  {\bf 94}, 081801 (2005);
%
  S.~Abe {\it et al.}  [KamLAND Collaboration],
  arXiv:0801.4589 [hep-ex].


\bibitem{atm}
  Y.~Ashie {\it et al.}  [Super-Kamiokande Collaboration],
  Phys.\ Rev.\ D {\bf 71}, 112005 (2005).

\bibitem{chooz}
M.~Apollonio {\it et al.},
Eur.\ Phys.\ J.\ C {\bf 27}, 331 (2003).

\bibitem{k2k}
E.~Aliu {\it et al.}  [K2K Collaboration],
  Phys.\ Rev.\ Lett.\  {\bf 94}, 081802 (2005). 

\bibitem{minos}
D. G. Michael {\it et al.}, [MINOS Collaboration],
  arXiv:hep-ex/0607088.

\bibitem{limits}
  M.~C.~Gonzalez-Garcia and M.~Maltoni,
  arXiv:0704.1800 [hep-ph];
%
M.~Maltoni, 
T.~Schwetz, M.~A.~Tortola and J.~W.~F.~Valle,
New J.\ Phys.\  {\bf 6}, 122 (2004), hep-ph/0405172 v5;
%
  S.~Choubey,
  arXiv:hep-ph/0509217;
%
  S.~Goswami,
  Int.\ J.\ Mod.\ Phys.\ A {\bf 21}, 1901 (2006);
%
  A.~Bandyopadhyay, 
S.~Choubey, S.~Goswami, S.~T.~Petcov and D.~P.~Roy,
  Phys.\ Lett.\ B {\bf 608}, 115 (2005);
%
  G.~L.~Fogli {\it et al.}, 
  Prog.\ Part.\ Nucl.\ Phys.\  {\bf 57}, 742 (2006).

\bibitem{golden}
  A.~Cervera, A.~Donini, M.~B.~Gavela, J.~J.~G\'{o}mez-Cadenas, P.~Hernandez, O.~Mena and S.~Rigolin,
  Nucl.\ Phys.\ B {\bf 579}, 17 (2000)
  [Erratum-ibid.\ B {\bf 593}, 731 (2001)].

\bibitem{icarus}
  P.~Aprili {\it et al.}  [ICARUS Collaboration],
CERN-SPSC-2002-027 (2002).


\bibitem{opera}
  M.~Guler {\it et al.}  [OPERA Collaboration],
CERN-SPSC-2000-028 (2000);


\bibitem{t2k}
  Y.~Itow {\it et al.},
  arXiv:hep-ex/0106019.


\bibitem{nova}
  D.~S.~Ayres {\it et al.}  [NOvA Collaboration],
  arXiv:hep-ex/0503053.

\bibitem{chooz2}
  F.~Ardellier {\it et al.},
  arXiv:hep-ex/0405032;
%
  F.~Ardellier {\it et al.}  [Double Chooz Collaboration],
  arXiv:hep-ex/0606025.

\bibitem{huber10}
  P.~Huber,
M.~Lindner, M.~Rolinec, T.~Schwetz and W.~Winter,
  Phys.\ Rev.\ D {\bf 70}, 073014 (2004)
and references therein.


\bibitem{zucc}
P.~Zucchelli,
Phys.\ Lett.\ B {\bf 532}, 166 (2002).

\bibitem{lindroos}
 M.~Lindroos,
  arXiv:physics/0312042;
%
  M.~Lindroos,
  Nucl.\ Phys.\ Proc.\ Suppl.\  {\bf 155}, 48 (2006).

\bibitem{betabeampage}
http://beta-beam.web.cern.ch/beta\%2Dbeam/


\bibitem{geer}
  S.~Geer,
  Phys.\ Rev.\  D {\bf 57}, 6989 (1998)
  [Erratum-ibid.\  D {\bf 59}, 039903 (1999)].

\bibitem{mind}
A. Cervera, Talk at NuFact07, Okayama University, Okayama, Japan, 
August 6-11, 2007, http://fphy.hep.okayama-u.ac.jp/nufact07/

\bibitem{crossthomas}
  P.~Huber, M.~Mezzetto and T.~Schwetz,
  arXiv:0711.2950 [hep-ph].

\bibitem{iss}
http://www.hep.ph.ic.ac.uk/iss/

\bibitem{issphysics}
  A.~Bandyopadhyay {\it et al.}  [ISS Physics Working Group],
  arXiv:0710.4947 [hep-ph].

\bibitem{intrinsic}
  J.~Burguet-Castell, 
M.~B.~Gavela, J.~J.~G\'{o}mez-Cadenas, P.~Hernandez and O.~Mena,
  Nucl.\ Phys.\ B {\bf 608}, 301 (2001).

\bibitem{minadeg}
  H.~Minakata and H.~Nunokawa,
  JHEP {\bf 0110}, 001 (2001).

\bibitem{th23octant}
  G.~L.~Fogli and E.~Lisi,
  Phys.\ Rev.\ D {\bf 54}, 3667 (1996).

\bibitem{eight}
  V.~Barger, D.~Marfatia and K.~Whisnant,
  Phys.\ Rev.\ D {\bf 65}, 073023 (2002).

\bibitem{diffLnE}
  H.~Minakata and H.~Nunokawa,
  Phys.\ Lett.\  B {\bf 413}, 369 (1997);
%
  V.~Barger, D.~Marfatia and K.~Whisnant,
  Phys.\ Rev.\  D {\bf 66}, 053007 (2002);
%
  V.~Barger, D.~Marfatia and K.~Whisnant,
  Phys.\ Lett.\  B {\bf 560}, 75 (2003);
%
  O.~Mena and S.~J.~Parke,
  Phys.\ Rev.\  D {\bf 70}, 093011 (2004);
%
  O.~Mena Requejo, S.~Palomares-Ruiz and S.~Pascoli,
  Phys.\ Rev.\  D {\bf 72}, 053002 (2005);
%
  M.~Ishitsuka, T.~Kajita, H.~Minakata and H.~Nunokawa,
  Phys.\ Rev.\  D {\bf 72}, 033003 (2005);
%
  K.~Hagiwara, N.~Okamura and K.~i.~Senda,
  Phys.\ Rev.\  D {\bf 76}, 093002 (2007).

\bibitem{t2ksimulation}
  P.~Huber, M.~Lindner and W.~Winter,
  Nucl.\ Phys.\  B {\bf 645}, 3 (2002);
%
  P.~Huber, M.~Lindner and W.~Winter,
  Nucl.\ Phys.\  B {\bf 654}, 3 (2003).

\bibitem{silver}
  A.~Donini, D.~Meloni and P.~Migliozzi,
  Nucl.\ Phys.\  B {\bf 646}, 321 (2002);
%
  D.~Autiero {\it et al.},
  Eur.\ Phys.\ J.\  C {\bf 33}, 243 (2004).

\bibitem{dissappear}
  A.~Donini, E.~Fernandez-Martinez and S.~Rigolin,
  Phys.\ Lett.\  B {\bf 621}, 276 (2005);
%
  A.~Donini, E.~Fernandez-Martinez, D.~Meloni and S.~Rigolin,
  Nucl.\ Phys.\  B {\bf 743}, 41 (2006).

\bibitem{pee}
  S.~K.~Agarwalla, S.~Choubey, S.~Goswami and A.~Raychaudhuri,
  Phys.\ Rev.\  D {\bf 75}, 097302 (2007).

\bibitem{addatm}
  P.~Huber, M.~Maltoni and T.~Schwetz,
  Phys.\ Rev.\  D {\bf 71}, 053006 (2005);

\bibitem{cernmemphys}
  J.~E.~Campagne, M.~Maltoni, M.~Mezzetto and T.~Schwetz,
  JHEP {\bf 0704}, 003 (2007).

\bibitem{addreact}
  P.~Huber, M.~Lindner, T.~Schwetz and W.~Winter,
  Nucl.\ Phys.\  B {\bf 665}, 487 (2003).

\bibitem{magic}
  P.~Huber and W.~Winter,
  Phys.\ Rev.\ D {\bf 68}, 037301 (2003).

\bibitem{magic2}
  A.~Y.~Smirnov,
  arXiv:hep-ph/0610198.

\bibitem{petcov}
  M.~Freund, M.~Lindner, S.~T.~Petcov and A.~Romanino,
  Nucl.\ Phys.\  B {\bf 578}, 27 (2000).

\bibitem{nufactoptim}
  P.~Huber, M.~Lindner, M.~Rolinec and W.~Winter,
  Phys.\ Rev.\  D {\bf 74}, 073003 (2006).

\bibitem{paper1}
  S.~K.~Agarwalla, A.~Raychaudhuri and A.~Samanta,
  Phys.\ Lett.\ B {\bf 629}, 33 (2005).

\bibitem{betaino1}
  S.~K.~Agarwalla, S.~Choubey and A.~Raychaudhuri,
  Nucl.\ Phys.\  B {\bf 771}, 1 (2007).

\bibitem{betaino2}
  S.~K.~Agarwalla, S.~Choubey and A.~Raychaudhuri,
  Nucl.\ Phys.\  B {\bf 798}, 124 (2008).

\bibitem{oldpapers}
  M.~Mezzetto,
  J.\ Phys.\ G {\bf 29}, 1771 (2003)
  [arXiv:hep-ex/0302007].
%
  M.~Mezzetto,
  Nucl.\ Phys.\ Proc.\ Suppl.\  {\bf 143}, 309 (2005).
%
  M.~Mezzetto,
  Nucl.\ Phys.\ Proc.\ Suppl.\  {\bf 155}, 214 (2006);
%

\bibitem{donini130}
  A.~Donini, E.~Fernandez-Martinez, P.~Migliozzi, S.~Rigolin and 
L.~Scotto Lavina,
  Nucl.\ Phys.\  B {\bf 710}, 402 (2005).


\bibitem{doninibeta}
  A.~Donini, E.~Fernandez, P.~Migliozzi, S.~Rigolin, L.~Scotto Lavina, 
  T.~Tabarelli de Fatis and F.~Terranova,
  arXiv:hep-ph/0511134;
%
  A.~Donini, E.~Fernandez-Martinez, P.~Migliozzi, S.~Rigolin, 
  L.~Scotto Lavina, T.~Tabarelli de Fatis and F.~Terranova,
  Eur.\ Phys.\ J.\  C {\bf 48}, 787 (2006).

\bibitem{newdonini}
  P.~Coloma, A.~Donini, E.~Fernandez-Martinez and J.~Lopez-Pavon,
  arXiv:0712.0796 [hep-ph].

\bibitem{bc}
  J.~Burguet-Castell, D.~Casper, E.~Couce, J.~J.~G\'{o}mez-Cadenas and P.~Hernandez,
  Nucl.\ Phys.\  B {\bf 725}, 306 (2005).
%

\bibitem{bc2}
  J.~Burguet-Castell, D.~Casper, J.~J.~G\'{o}mez-Cadenas, P.~Hernandez and F.~Sanchez,
  Nucl.\ Phys.\  B {\bf 695}, 217 (2004).

\bibitem{fnal}
  A.~Jansson, O.~Mena, S.~Parke and N.~Saoulidou,
  arXiv:0711.1075 [hep-ph].

\bibitem{betaoptim}
  P.~Huber, M.~Lindner, M.~Rolinec and W.~Winter,
  Phys.\ Rev.\  D {\bf 73}, 053002 (2006).

\bibitem{volpe}
C.~Volpe,
  J.\ Phys.\ G {\bf 34}, R1 (2007).

\bibitem{doninialter}
  A.~Donini and E.~Fernandez-Martinez,
  Phys.\ Lett.\ B {\bf 641}, 432 (2006).

\bibitem{rparity}
  R.~Adhikari, S.~K.~Agarwalla and A.~Raychaudhuri,
  Phys.\ Lett.\ B {\bf 642}, 111 (2006).
%
  S.~K.~Agarwalla, S.~Rakshit and A.~Raychaudhuri,
  Phys.\ Lett.\  B {\bf 647}, 380 (2007).

\bibitem{boulby}
  D.~Meloni, O.~Mena, C.~Orme, S.~Palomares-Ruiz and S.~Pascoli,
  arXiv:0802.0255 [hep-ph].

\bibitem{rubbia}
    C.~Rubbia, A.~Ferrari, Y.~Kadi and V.~Vlachoudis,
  Nucl.\ Instrum.\ Meth.\ A {\bf 568}, 475 (2006);
  C.~Rubbia,
  arXiv:hep-ph/0609235.

\bibitem{mori}
  Y.~Mori,
  Nucl.\ Instrum.\ Meth.\  A {\bf 562}, 591 (2006).


\bibitem{ino}
  M.~S.~Athar {\it et al.}  [INO Collaboration],
 A Report of the INO Feasibility Study,\\
{http://www.imsc.res.in/~ino/OpenReports/INOReport.pdf}

\bibitem{volpelow}
  C.~Volpe,
  J.\ Phys.\ G {\bf 30}, L1 (2004). 

\bibitem{nf07sc}
S. Choubey, Talk at NuFact07, Okayama University, Okayama, Japan, 
August 6-11, 2007, http://fphy.hep.okayama-u.ac.jp/nufact07/;
  S.~K.~Agarwalla, S.~Choubey and A.~Raychaudhuri,
  arXiv:0712.4072 [hep-ph].

\bibitem{beamnorm} 
  B.~Autin, R.~C.~Fernow, S.~Machida and D.~A.~Harris,
  J.\ Phys.\ G {\bf 29}, 1637 (2003);
%
  F.~Terranova, A.~Marotta, P.~Migliozzi and M.~Spinetti,
  Eur.\ Phys.\ J.\  C {\bf 38}, 69 (2004).

\bibitem{beta}
L. P. Ekstrom and R. B. Firestone, WWW Table of Radioactive Isotopes, \\
database version 2/28/99 from URL http://ie.lbl.gov/toi/

\bibitem{mats_talk_RAL_2008}
M. Lindroos, 
Talk at the ``First plenary meeting of the 
International Design Study for the Neutrino Factory'', 
RAL, United Kingdom, 16-17 January 2008. 

\bibitem{uno}
  C.~K.~Jung,
  AIP Conf.\ Proc.\  {\bf 533}, 29 (2000)


\bibitem{hk}
  Y.~Itow {\it et al.},
  arXiv:hep-ex/0106019.


\bibitem{memp}
  A.~de Bellefon {\it et al.},
  arXiv:hep-ex/0607026.

\bibitem{msw1}
  L.~Wolfenstein,
  Phys.\ Rev.\ D {\bf 17}, 2369 (1978);

\bibitem{msw2}
  S.~P.~Mikheev and A.~Y.~Smirnov,
  Sov.\ J.\ Nucl.\ Phys.\  {\bf 42}, 913 (1985)
  [Yad.\ Fiz.\  {\bf 42}, 1441 (1985)];
%
  S.~P.~Mikheev and A.~Y.~Smirnov,
  Nuovo Cim.\ C {\bf 9}, 17 (1986).

\bibitem{msw3}
  V.~D.~Barger, K.~Whisnant, S.~Pakvasa and R.~J.~N.~Phillips,
  Phys.\ Rev.\ D {\bf 22}, 2718 (1980).

\bibitem{freund}
  M.~Freund, P.~Huber and M.~Lindner,
  Nucl.\ Phys.\  B {\bf 615}, 331 (2001).

\bibitem{prem}
  A.~M.~Dziewonski and D.~L.~Anderson,
  Phys.\ Earth Planet.\ Interiors {\bf 25}, 297 (1981);
\\
S.~V.~Panasyuk, Reference Earth Model (REM) webpage,\\
 http://cfauves5.harvrd.edu/lana/rem/index.html.


\bibitem{solarprecision}
See for example,  
A.~Bandyopadhyay, S.~Choubey, S.~Goswami and S.~T.~Petcov,
  Phys.\ Rev.\ D {\bf 72}, 033013 (2005);
%
  J.~N.~Bahcall and C.~Pena-Garay,
  JHEP {\bf 0311}, 004 (2003).

\bibitem{tomography}
R.~J.~Geller and T.~Hara,
Nucl.\ Instrum.\ Meth.\  A {\bf 503}, 187 (2001).

\bibitem{globes}
  P.~Huber, J.~Kopp, M.~Lindner, M.~Rolinec and W.~Winter,
  Comput.\ Phys.\ Commun.\  {\bf 177}, 432 (2007);
%
  P.~Huber, M.~Lindner and W.~Winter,
  Comput.\ Phys.\ Commun.\  {\bf 167}, 195 (2005).

\bibitem{matsmoriond} M. Lindroos,
Talk at Moriond 
Workshop on ``Radioactive beams for nuclear physics
and neutrino physics"
March 17-22nd, 2003; http://moriond.in2p3.fr/radio/index.html.

\bibitem{optim2}
S.~K.~Agarwalla, S. Choubey, P. Huber, A. Raychaudhuri, W. Winter, 
in preparation.

\end{thebibliography}
\end{document}